Université du Québec
Institut National de la Recherche Scientifique
Centre Énergie Matériaux Télécommunications

# Optimal Energy Management for SmartGrids Considering Thermal Load and Dynamic Pricing

## Duong Tung Nguyen

A thesis presented to INRS-EMT
In partial fulfillment of the requirements for the degree of
Master of Science in Telecommunications

### Thesis Committee

| | |
|---|---|
| Chair of committee and internal committee member | Martin Maier Professor, INRS-EMT |
| External committee member | Francois Bouffard Assistant Professor, McGill University |
| Supervisor | Long Bao Le Assistant Professor, INRS-EMT |

# Résumé

Une participation plus active de la demande dans le fonctionnement du réseau et l'intégration efficace des ressources énergétiques distribuées (RED) telles que les véhicules électriques (VE), le stockage de l'énergie (SE), et les sources d'énergie renouvelables (SER) dans les systèmes d'alimentation existants sont des objectifs importants dans la conception du futur réseau électrique intelligent. En outre, les gestionnaires de réseau d'électricité ont besoin d'équilibrer l'offre et la demande d'électricité en temps réel pour maintenir la stabilité du réseau. Traditionnellement, les réseaux d'électricité sont construits avec une capacité de réserve pour répondre à la pointe de consommation électrique. Ici, la gestion de la pointe de consommation est essentielle dans la planification de l'extension du réseau tout en gardant le prix de l'électricité raisonnable. La maitrise de la demande en énergie (MDE) a été considérée comme un moyen efficace pour gérer la demande croissante en électricité. En particulier, MDE aide non seulement les services publics et les gestionnaires de réseau à différer la nécessité de mettre à niveau leurs réseaux d'électricité et d'améliorer le service à la clientèle, mais aussi à fournir aux clients la possibilité de réduire leurs factures d'électricité.

Un système de chauffage, de ventilation et de climatisation (CVC) est l'un des éléments clés dans la maitrise de la demande en énergie. En fait, les utilisateurs d'électricité désirent généralement maintenir la température intérieure à une valeur consigne optimale, qui dépend de leurs préférences et de l'état d'occupation du bâtiment. Toutefois, les utilisateurs peuvent accepter un petit écart entre la température de l'intérieur et le point de consigne désiré. Évidemment, ils sont plus confortables lorsque la température intérieure est plus proche du point de consigne préféré . Si la température souhaitable et la déviation de la température maximale admissible sont connues alors, la température de l'intérieur doit être maintenue pour qu'elle soit dans la plage de température requise. Ensuite, la consommation d'énergie de CVC peut être programmée d'une façon intelligente afin de réaliser des économies de coûts de l'électricité sans violer les exigences de confort des utilisateurs. Plus précisément, le système de CVC peut consommer plus de puissance pendant les heures de faible prix d'électricité pour prérefroidir (préchauffage) les bâtiments en été (hiver) alors qu'il peut réduire la consommation d'énergie pendant les heures où le prix de l'électricité est élevé, tout en maintenant la température à l'intérieur de la zone de confort grâce à l'inertie thermique du bâtiment.

La planification intelligente de l'énergie pour les systèmes de CVC est un sujet de recherche important pour plusieurs raisons. Tout d'abord, la charge de CVC contribue à une partie significative de la consommation totale d'énergie dans les bâtiments résidentiels et commerciaux. Par conséquent, il représente une part importante des factures d'électricité des utilisateurs. Ensuite, la consommation d'énergie agrégée des systèmes de CVC est la principale cause de la pointe de consommation électrique en été et en hiver. Enfin, tout en étant l'un des appareils les plus consommant d'énergie, le système de CVC offre une grande flexibilité dans le contrôle de sa consommation tout en respectant les besoins de confort des utilisateurs.

L'objectif général de cette thèse est d'étudier le problème de coordination de la planification énergétique des systèmes de CVC et les différentes ressources énergétiques distribuées afin de maximiser les avantages pour les utilisateurs. Les avantages de la coordination sont évalués dans deux scénarios d'application différents en utilisant des modèles mathématiques rigoureux. En outre, les caractéristiques de fonctionnement des différents RED, le concept de microréseau (MR), et les principes de négoce d'énergie dans le marché concurrentiel de l'électricité sont également étudiés. Le contenu de cette thèse est organisé comme suit. Le chapitre 2 couvre le fond pertinent, y compris la modélisation de la dynamique thermique du bâtiment et les techniques basiques d'optimisation.

Le chapitre 3 présente le cadre d'optimisation conjointe pour les véhicules électriques et les systèmes de CVC. Le chapitre 4 étudie un modèle d'optimisation stochastique pour la planification énergétique et les activités d'appels d'offres des microréseaux où la charge de CVC est utilisée pour compenser les incertitudes dans le processus de prise de décision de l'agrégateur du MR. Enfin, le chapitre 5 conclut la thèse en fournissant un bref résumé du travail accompli et énumérant quelques orientations futures de la recherche.

## 0.1 Un Modèle de la Dynamique Thermique pour les Bâtiments

En général, la température à l'intérieur peut être exprimée en fonction des caractéristiques thermiques du logement (la résistance thermique, la capacité thermique, la zone de la fenêtre), les conditions météorologiques (la température ambiante, le rayonnement solaire, la vitesse du vent, l'humidité), les gains internes (les occupants, la cuisine, le réfrigérateur, etc), et la puissance d'entrée de CVC. Avec quelques hypothèses simplificatrices, un modèle espace-état linéaire à temps discret représentant la dynamique thermique de la construction peut être calculé comme suit:

$$
\begin{aligned}
T_{j,\tau+1} &= A_j^{\mathrm{d}} T_{j,\tau} + B_j^{\mathrm{d}} U_{j,\tau} \\
T_{j,\tau}^{\mathrm{in}} &= C_j^{\mathrm{d}} T_{j,\tau}
\end{aligned}
\tag{1}
$$

où $T_{j,\tau} = [T_{j,\tau}^{\mathrm{in}},\ T_{j,\tau}^{\mathrm{m}},\ T_{j,\tau}^{\mathrm{e}}]'$ est le vecteur d'état et $U_{j,\tau} = [T_\tau^{\mathrm{a}},\ \Phi_\tau,\ \sigma_j \eta_j P_{j,\tau}^{\mathrm{hvac}}]'$ est le vecteur d'entrée du système. Les variables concernées sont: $T_{j,\tau}^{\mathrm{in}}$ est la température intérieure du bâtiment $j$ au temps $\tau$ et $P_{j,\tau}^{\mathrm{hvac}}$ est la puissance d'entrée du système de CVC dans le bâtiment $j$ au temps $\tau$. En plus, $T_\tau^{\mathrm{a}}$ est la température ambiante à l'instant $\tau$, $\Phi_\tau$ est l'irradiation solaire au temps $\tau$, $\sigma_j$ est l'indicateur de refroidissement/chauffage, et $\eta_j$ est le coefficient de performance de système CVC dans le bâtiment $j$. Les matrices $A_j^{\mathrm{d}}$ et $B_j^{\mathrm{d}}$ peuvent être calculées sur la base de la résistance thermique et les paramètres de la capacité thermique de la construction $j$. La matrice $A_j^{\mathrm{d}}$ représente le comportement dynamique du système, et la matrice $B_j^{\mathrm{d}}$ saisit l'impact des éléments d'entrée (la température ambiante, le rayonnement solaire et la puissance d'entrée de CVC) sur le comportement du système. $C_j^{\mathrm{d}} = [\ 1\ \ 0\ \ 0\ ]$.

Le reste du chapitre 3 porte sur les concepts de base de l'optimisation, y compris la formulation générale d'un problème d'optimisation mathématique, le concept de la programmation linéaire et la programmation stochastique en mettant l'accent sur un scénario basé sur l'optimisation stochastique en deux étapes. Enfin, certaines logiciels d'optimisation sont introduites.

## 0.2 Optimisation de la Planification Conjointe pour les Systèmes de CVC et les Véhicules Électriques

Le problème de coordination de la planification de l'énergie de systèmes de CVC et les véhicules électriques est étudié dans le chapitre 4. Les problèmes de planification de la consommation d'énergie de la maison et le chargement de VE sont souvent traités séparément dans la littérature. Pour remédier à cette limitation, un modèle d'optimisation unifié est proposé pour optimiser conjointement la planification des véhicules électriques et les systèmes CVC. En particulier, les véhicules électriques offrent une installation de stockage dynamique pour fournir de l'énergie pour les bâtiments résidentiels pendant les heures de pointe où l'énergie peut être transférée des véhicules électriques pour charger autres véhicules et fournir de l'énergie pour les systèmes de CVC. Le modèle proposé vise à réduire le coût total de l'électricité compte tenu le confort de l'utilisateur, l'occupation de la maison, les modes de déplacement des VE, la dynamique thermique, la demande d'électricité par les VE, et d'autres contraintes d'exploitation. Au moment de la rédaction du document, aucun des travaux antérieurs n'a examiné la conception détaillée et l'optimisation conjointe des véhicules électriques et la gestion d'énergie du bâtiment pour le meilleur de nos connaissances. Les principales contributions de cette partie de cette thèse peuvent être résumées comme suit :

- Un modèle d'optimisation global est proposé pour optimiser l'ordonnancement du VE et le CVC dans un quartier résidentiel. La formulation élaborée vise à atteindre un compromis flexible entre la minimisation du coût total de l'électricité et le maintien de préférence du confort de l'utilisateur. Le modèle tient compte des caractéristiques du système de CVC, la dynamique thermique, la préférence de confort climatique de l'utilisateur, le modèle d'état de la batterie, les habitudes de déplacement de l'utilisateur, et les modes d'occupation des ménages. Des extensions possibles du cadre proposé pour capturer les divers facteurs de modélisation d'incertitude sont également discutées.

- Les impacts des différents paramètres de conception et du système, qui contrôlent le coût de l'électricité et le confort de l'utilisateur, sur le fonctionnement et la performance du système ainsi que les avantages économiques de l'application du cadre de contrôle proposé par rapport à un système de contrôle non optimisé pour un scénario d'une seule maison, sont présentés

- Les avantages de l'application du modèle de contrôle proposé pour le scénario de plusieurs maisons par rapport au cas où chaque ménage optimise sa consommation d'énergie séparément sont illustrés par résultats numériques. Plus précisément, l'optimisation de la planification des VE et de l'énergie pour plusieurs maisons dans un quartier résidentiel permet de diminuer les coûts de l'électricité et de réduire la demande élevée de puissance pendant les heures de pointe.

Nous considérons l'interaction entre les véhicules électriques et les systèmes de CVC dans un quartier résidentiel qui se compose d'un certain nombre de ménages. Chaque foyer est censé avoir un système de CVC et peut avoir un nombre arbitraire de véhicules électriques. Nous supposons qu'il existe un agrégateur qui recueille toutes les informations nécessaires des véhicules électriques et des systèmes de CVC de tous les ménages, pour prendre des décisions de contrôle. Nous considérons un modèle fendu de temps où il y a NH intervalles de temps dans la période d'optimisation (par exemple, 24 heures pour une période d'optimisation d'une journée) et les décisions de planification de l'énergie sont faites pour chaque intervalle de temps. Basé sur les informations de prix et la demande d'électricité, chaque agrégateur décide combien d'énergie il doit importer du réseau à chaque intervalle de temps, comment la répartir et planifier son utilisation et l'échange entre ses composantes, y compris les CVC et les véhicules électriques. Nous supposons que les véhicules électriques ne peuvent être chargés ou déchargés que s'ils sont stationnés à la maison.

La formulation mathématique commune du rechargement de VE et le problème de gestion de l'énergie de la maison pour la minimisation des coûts, est présentée ci-après. Tout d'abord, la puissance électrique totale importée de la grille à l'intervalle de temps t peut être écrite comme suit.

$$P_t^{\text{grid}} = \sum_{j=1}^{NB} P_{j,t}^{\text{hvac}} + \sum_{j=1}^{NB} \sum_{e=1}^{E_k} \left[ P_{j,e,t}^{\text{ev,c}} - P_{j,e,t}^{\text{ev,d}} \right], \forall t \qquad (2)$$

qui est égale à la puissance de recharge de VE, plus la consommation d'énergie du système de CVC moins la puissance de décharge de VE additionnées sur tous les ménages. L'objectif et les contraintes du problème d'optimisation sous-jacent sont décrits ci-après.

**La Fonction Objectif**

La fonction objectif, qui est la somme du coût de l'électricité et le coût de l'inconfort, peut être écrite de la manière suivante.

$$J_{\text{tot}} = \sum_{i=1}^{N} \left( \sum_{k=1}^{M} P_{k,i}^{\text{hvac}} + \sum_{k=1}^{M} \sum_{j=1}^{J_k} \left[ P_{k,j,i}^{\text{ev},c} - P_{k,j,i}^{\text{ev},d} \right] \right) e_i \Delta T$$
$$+ \sum_{i=1}^{N} \sum_{k=1}^{M} w_k a_{k,i+1} |T_{k,i+1}^{\text{in}} - T_{k,i+1}^{\text{d}}|$$

(3)

Nous sommes maintenant prêts à décrire l'ensemble des contraintes pour le problème d'optimisation considéré.

**Les Contraintes Thermiques**

Le modèle thermique dynamique de temps discret représente une contrainte du problème d'optimisation.

$$T_{j,t+1} = A_j^{\text{d}} T_{j,t} + B_j^{\text{d}} U_{j,t}, \quad \forall j, \ t$$
$$T_{j,t}^{\text{in}} = C_j^{\text{d}} T_{j,t}, \quad \forall j, \ t$$

(4)

La température intérieure doit être maintenue dans la plage de confort.

$$a_{j,t} |T_{j,t}^{\text{in}} - T_{j,t}^{\text{d}}| \leq \delta_{j,t},$$

(5)

for $j = 1, 2, \ldots, NB$ and $t = 2, \ldots, NH + 1$.

En outre, la consommation d'énergie d'un système de CVC est limitée par sa puissance nominale.

$$0 \leq P_{j,t}^{\text{hvac}} \leq P_j^{\text{hvac,max}},$$

(6)

for $j = 1, 2, \ldots, NB$ and $t = 1, \ldots, NH$.

## SOC et les Contraintes de Puissance de Chargement

$$SOC_{j,e,t+1} = SOC_{j,e,t} + \frac{\eta_{j,e}^c P_{j,e,t}^{ev,c} \Delta T}{E_{j,e}^{cap}} - \frac{P_{j,e,t}^{ev,d} \Delta T}{\eta_{j,e}^d E_{j,e}^{cap}}, \text{ if } t \notin [t_{j,e,l}^{(1)}, t_{j,e,l}^{(2)}], \quad \forall j, t, e, l \quad (7)$$

$$SOC_{j,e,t+\Lambda_{j,e,l}} = SOC_{j,e,i} - \frac{d_{j,e,l} * m_j}{E_{j,e}^{cap}}, \text{ if } t = t_{j,e,l}^{(1)}, \quad \forall j, t, e, l \quad (8)$$

$$SOC_{j,e,t_{j,e,l}^{(2)}} \leq SOC_{j,e,t} \leq SOC_{j,e,t_{j,e,l}^{(1)}}, \text{ if } t \in [t_{j,e,l}^{(1)}, t_{j,e,l}^{(2)}], \quad \forall j, t, e, l. \quad (9)$$

$$SOC_{j,e}^{min} \leq SOC_{j,e,t} \leq SOC_{j,e}^{max}, \quad \forall j, e, t \quad (10)$$

$$0 \leq P_{j,e,t}^{ev,c} \leq b_{j,e,t} P_{j,e}^{ev,c,max}, \quad \forall j, e, t \quad (11)$$

$$0 \leq P_{j,e,t}^{ev,d} \leq b_{j,e,t} P_{j,e}^{ev,d,max}, \quad \forall j, e, t. \quad (12)$$

## Les Contraintes du Réseau

S'il n'est pas autorisé de vendre l'énergie de déchargement des VE au réseau principal, nous avons:

$$0 \leq P_t^{grid} \leq P_t^{max}, \quad \forall t \qquad (13)$$

où $P_t^{max}$ est la puissance maximale qui peut être importé du réseau. Par contre, si le service de vente de l'énergie de déchargement des VE est autorisé, nous avons la contrainte suivante

$$-P_t^{max} \leq P_t^{grid} \leq P_t^{max}, \quad \forall t \qquad (14)$$

Par souci de simplicité, nous supposons que le prix de revente de l'électricité est égal à son prix d'achat. Le problème d'optimisation conjointe décrit ci-dessus est en effet un programme linéaire à grande échelle. Nous utilisons le logiciel CVX pour résoudre le problème d'optimisation proposée. L'efficacité de l'approche proposée est démontrée par de nombreux résultats numériques. Plus précisément, les simulations ont été effectuées dans deux scénarios : un seul ménage et plusieurs ménages (ou une communauté). Les résultats numériques ont montré que la performance de l'optimal proposé est fortement influencé par les paramètres du système, y compris les habitudes de déplacement des véhicules électriques et les paramètres de confort de température d'utilisateurs (par exemple, le maximum écart admissible de la température et le coût de l'écart de température). En outre, V2G apporte plusieurs avantages considérables par rapport au cas où il n'est pas autorisé. Enfin, si plusieurs ménages dans une communauté optimisent conjointement la consommation énergétique de leurs véhicules électriques et des systèmes de CVC, le résultat surpasse de manière significative le résultat dans le cas où chaque famille essaie d'optimiser la consommation d'énergie de leur véhicule électrique et système CVC indépendamment et séparément.

## 0.3 Soumission Optimale du Microréseau d'un Bâtiment

Chapitre 5 étudie un scénario d'application différent où la charge flexible de CVC a été utilisée pour compenser les incertitudes du jour d'après de la décision d'appel d'offres d'énergie de l'agrégateur MR. Plus précisément, nous avons examiné non seulement la variation du prix de l'électricité dans le temps, mais aussi la variation de la production d'énergie renouvelable dans le temps dans la programmation de la consommation d'énergie de CVC. Pour le meilleur de notre connaissance, il n'y a pas un travail existant qui propose un tel cadre d'appel d'offres d'énergie économique. Notre conception vise à exploiter le potentiel de l'utilisation de systèmes de CVC et la charge thermique associée pour faire face aux incertitudes de SER et de maximiser le bénéfice du MR dans le marché de l'électricité. Les principales contributions de cette partie de la thèse peuvent être résumées comme suit :

- Nous proposons un modèle global sur lequel on se base pour développer une stratégie optimale d'ordonnancement le jour d'après pour un MR dans un marché concurrentiel d'électricité. Le modèle proposé vise un équilibre entre la maximisation des revenus du MR et une minimisation de la charge, du déversement de l'énergie renouvelable ainsi que la déviation de la soumission, tout en maintenant les exigences de confort des utilisateurs et d'autres contraintes du système. Le modèle proposé est nouveau par le fait qu'il nous permet d'exploiter les propriétés thermiques dynamique des bâtiments pour compenser la variabilité de la production d'énergie renouvelable, ce qui peut améliorer de manière significative le bénéfice du MR. Plus précisément, la charge de CVC est utilisée comme une source de la réponse aux demandes et intégrée dans la stratégie de soumission optimale de l'agrégateur de MR dans le marché de l'électricité.

- Le modèle d'optimisation est formulé comme un problème de programmation stochastique en deux étapes où les incertitudes sont capturées à l'aide de la méthode de simulation de Monte-Carlo. Les variables des première et deuxième étapes sont définies d'une manière appropriée pour assurer l'efficacité des opérations des systèmes d'alimentation intégrants les différentes sources d'énergie renouvelables

- Nous présentons des nombreux résultats de simulation pour démontrer les avantages de la coordination des opérations des systèmes de CVC et des ressources énergétiques renouvelables par rapport au cas non coordonné où les systèmes de CVC et les autres composants du MR visent à optimiser leur consommation d'énergie/production indépendamment. La performance du système proposé est également comparée à celle sous l'exigence stricte de confort climatique où aucun écart de température n'est autorisé. Enfin, une analyse de sensibilité est effectuée pour évaluer les effets des différents paramètres de système et de conception sur la solution optimale.

Nous considérons un MR à grande échelle qui se compose de plusieurs unités de production d'énergie renouvelable, des unités classiques, d'un certain nombre de bâtiments avec des charges connexes, ainsi qu'une installation de stockage de la batterie optionnelle. Les unités de production d'énergie renouvelable comprennent des panneaux solaires et des éoliennes. Dans cette étude, les unités de production classiques (DG) se réfèrent aux unités de production d'énergie non renouvelable telles que des microturbines, des piles à combustible, et des générateurs diesel. En général, l'agrégateur MR désire maximiser l'utilisation de la production d'énergie renouvelable. La quantité réduite et excessive d'énergie nécessaire pour équilibrer la charge locale peut être autorisée que parle commerce avec la grille principale à travers le point de couplage commun (PCC) ou en exécutant des unités conventionnelles. L'agrégateur de MR prendra des décisions sur l'achat d'électricité du marché ou d'exécution des unités conventionnelles locales en fonction de divers facteurs tels que le prix de l'électricité, les états des unités conventionnelles, et le coût marginal de fonctionnement des unités conventionnelles. Nous considérons la planification de l'énergie et le problème d'appel d'offres pour le MR dans des intervalles de temps discrets, noté t dans notre modèle, pendant une période de planification de NH intervalles de temps.

Le MR est censé être un preneur de prix dans le marché de l'électricité. Pendant les intervalles de temps où la production locale d'énergie est excédentaire, le MR vendrait sa puissance au réseau principal. En face, si la production locale ne suffit pas à répondre à sa charge locale, le MR aurait besoin d'acheter de l'électricité à partir de la grille principale. Chaque jour, le MR doit présenter ses soumissions horaires sur le marché du jour d'après (day-ahead) plusieurs heures avant la livraison physique de l'énergie. Les offres de MR comprennent à la fois la vente et l'achat d'offres d'électricité. En plus, le MR peut participer au marché de l'électricité en temps réel pour compléter toute déviation d'alimentation de l'horaire du jour précédent. Trouver une stratégie de soumission horaire optimale pour le MR est une tâche difficile à cause de diverses incertitudes dans le système, ce qui peut entraîner un écart significatif entre la livraison de puissance prévue et la livraison de puissance en temps réel. Si le MR ne peut pas suivre la puissance prévue un jour d'après, une pénalité sera appliquée à l'écart de la soumission. Lorsque la production d'électricité renouvelable est plus élevée que prévu, il est parfois préférable de limiter le surplus d'énergie renouvelable afin d'éviter une pénalité élevée sur la déviation de la soumission. La capacité de stockage thermique des bâtiments peut aider à atténuer les effets des incertitudes d'énergie renouvelable, en augmentant la consommation d'énergie de CVC lorsque la production d'énergie renouvelable est plus élevée que prévu, et vice versa. En exploitant cet aspect de planification de la puissance de CVC, il est donc prévu que la livraison de puissance en temps réel sera plus proche de celle du jour d'après, et le déversement de l'énergie renouvelable est réduit.

Il existe différentes sources d'incertitudes dans notre modèle proposé, qui résultent de la production d'énergie renouvelable, la charge totale des systèmes non-CVC, la température ambiante, et les prix de l'électricité en temps réel et du jour d'après. La simulation de Monte-Carlo est utilisée pour générer des scénarios qui représentent ces paramètres d'incertitude sur la base des fonctions de distribution correspondantes. Une technique de réduction de scénario est utilisée pour réduire la charge de calcul. Pour plus de simplicité, la vitesse du vent, la charge des non-CVC, le rayonnement solaire, la température ambiante, les prix de l'électricité en temps réel et du jour d'après sont supposés suivre une distribution normale où les moyennes sont mises égales aux valeurs de prévisions et les écarts-types des valeurs moyennes sont respectivement 10%, 3%, 10%, 5%, 5%, et 15%.

Le problème d'ordonnancement d'énergie et de l'offre est formulé comme un programme stochastique en deux étapes. Les entrées du problème sous-jacent comprennent le scénario Monte-Carlo qui représente les incertitudes du système. Les sorties du problème d'optimisation sont constituées par les ensembles de décisions des première et deuxième étapes. Dans cette thèse, les décisions de la première étape comprennent les états d'engagement de toutes les unités classiques et les quantités de soumission horaires proposées dans le marché du jour d'après (day-ahead). Les décisions de la deuxième étape comprennent l'expédition d'alimentation de toutes les unités de production, la quantité de déversement involontaire de la charge, la livraison de puissance en temps réel entre le MR et le réseau principal, et les décisions de charge/décharge de la batterie.

La fonction objectif et les contraintes dans le problème sous-jacent d'optimisation stochastique en deux étapes sont décrites dans ce qui suit. Toutes les variables de décision de deuxième étape sont désignées par le l'exposant $s$ qui représente le scénario $s$.

**La Fonction Objectif**

Notre objectif de conception est de maximiser la fonction objectif suivante.

$$
\begin{aligned}
\max \quad & -\sum_{i=1}^{NG}\sum_{t=1}^{NH}(SU_{i,t}+SD_{i,t}) - \sum_{s=1}^{NS}\rho_s\sum_{i=1}^{NG}\sum_{t=1}^{NH}C(P_{i,t}^s) \\
& -\sum_{s=1}^{NS}\rho_s\sum_{t=1}^{NH}\pi_{j,t}\sum_{j=1}^{NB}|T_{j,t+1}^s - T_{j,t+1}^d| \\
+\sum_{s=1}^{NS}\rho_s\sum_{t=1}^{NH}\Delta T & \left\{ P_t e_t^{s,\mathsf{DA}} + (P_t^s - P_t)e_t^{s,\mathsf{RT}} - \psi_t|P_t^s - P_t| \right. \\
& -C_k^{\mathsf{deg}}\sum_{k=1}^{NK}(\frac{P_{k,t}^{s,\mathsf{d}}}{\eta_k^d} + \eta_k^c P_{k,t}^{s,\mathsf{c}}) - V^{LL}LS_t^s \\
& \left. -V_t^{\mathsf{W}}\cdot\sum_{w=1}^{NW}P_{w,t}^{s,\mathsf{ws}} - V_t^{\mathsf{PV}}\cdot\sum_{p=1}^{NP}P_{p,t}^{s,\mathsf{pvs}} \right\}
\end{aligned}
\tag{15}
$$

où $P_t$ est la soumission horaire que le MR présente sur le marché jour d'après, $P_t^s$ est la livraison de puissance en temps réel. Le décalage entre la puissance jour d'après prévue et la livraison de puissance en temps réel $|P_t^s - P_t|$ est en effet la quantité d'énergie que le MR échange dans le marché d'équilibrage en temps réel. La fonction objectif proposée représente le bénéfice attendu du MR qui est égal au chiffre d'affaires prévu atteindre par la négociation à la fois au jour d'après et à l'équilibre des marchés, moins le coût d'exploitation du MR. Le coût d'exploitation MR est composé du coût de démarrage, du coût de l'arrêt, du coût d'exploitation des unités conventionnelles, et d'autres coûts, y compris des coûts de la température d'inconfort des utilisateurs, du coût de dégradation de la batterie, des pénalités dues au vent/réduction de l'énergie solaire, et déversement involontaire de la charge.

D'autres contraintes du problème d'optimisation sont décrites dans ce qui suit.

### Équilibre de Puissance

Pour chaque scénario $s$, la somme de la production totale d'électricité de toutes les unités locales de production, le montant de déversement involontaire de charge, et la puissance de charge/décharge des unités de la batterie doit être égale à la somme de la livraison de puissance en temps réel et les charges de systèmes CVC et non-CVC. Notez que $P_{w,t}^s$ et $P_{w,t}^{s,\text{ws}}$ sont respectivement la puissance maximale disponible du vent et la quantité de déversement d'énergie éolienne à l'instant $t$ dans le scénario $s$. La différence entre eux est la production d'énergie éolienne réelle à l'instant $t$ dans le scénario $s$. L'explication similaire est appliquée à la production d'énergie solaire. L'équation d'équilibre de puissance pour chaque temps $t$ et le scénario $s$ est donnée comme suit:

$$\sum_{i=1}^{NG} P_{i,t}^s + \sum_{w=1}^{NW}(P_{w,t}^s - P_{w,t}^{s,\text{ws}}) + \sum_{p=1}^{NP}(P_{p,t}^s - P_{p,t}^{s,\text{pvs}}) + LS_t^s + \sum_{k=1}^{NK}(P_{k,t}^{s,\text{d}} - P_{k,t}^{s,\text{c}})$$
$$= P_t^s + \sum_{j=1}^{NB} P_{j,t}^{s,\text{hvac}} + L_t^s, \quad \forall\, s,\ t. \quad (16)$$

### Échange de Puissance avec le Réseau Principal

$$-P_t^{\text{g,max}} \le P_t^s \le P_t^{\text{g,max}}, \quad \forall\, s,\ t. \qquad (17)$$
$$-P_t^{\text{g,max}} \le P_t \le P_t^{\text{g,max}}, \quad \forall\, t. \qquad (18)$$

### Les Contraintes des Unités Conventionnelles

Le coût d'exploitation de l'unité conventionnelle $i$ peut être modélisé approximativement par une fonction linéaire par morceaux comme suit:

$$C(P_{i,t}^s) = a_i I_{i,t} + \Delta T \sum_{m=1}^{N_i} \lambda_{i,t}(m) P_{i,t}^s(m), \quad \forall\ i,\ t,\ s \tag{19}$$

$$0 \leq P_{i,t}^s(m) \leq P_{i,t}^{\mathsf{max}}(m), \quad \forall\ i,\ t,\ s \tag{20}$$

$$P_{i,t}^s = P_i^{\mathsf{min}} I_{i,t} + \sum_{m=1}^{N_i} P_{i,t}^s(m), \quad \forall\ i,\ t,\ s \tag{21}$$

Les contraintes suivantes représentent les limites de génération de puissance (22), le taux limite d'accélération/décélération (23)–(24), les délais minimum de ON/ OFF (25)–(26), et la relation entre les indicateurs de démarrage et d'arrêt (27)–(28) de l'unité conventionnelle $i$ .

$$P_i^{\mathsf{min}} I_{i,t} \leq P_{i,t}^s \leq P_i^{\mathsf{max}} I_{i,t}, \quad \forall\ i,\ t,\ s \tag{22}$$

$$P_{i,t}^s - P_{i,t-1}^s \leq UR_i(1 - y_{i,t}) + P_i^{\mathsf{min}} y_{i,t}, \quad \forall\ i,\ t,\ s \tag{23}$$

$$P_{i,t-1}^s - P_{i,t}^s \leq DR_i(1 - z_{i,t}) + P_i^{\mathsf{min}} z_{i,t}, \quad \forall\ i,\ t,\ s \tag{24}$$

$$\sum_{h=t}^{t+UT_i-1} I_{i,t} \geq UT_i y_{i,t}, \quad \forall\ i,\ t \tag{25}$$

$$\sum_{h=t}^{t+DT_i-1} (1 - I_{i,t}) \geq DT_k z_{i,t},, \quad \forall\ i,\ t \tag{26}$$

$$y_{i,t} - z_{i,t} = I_{i,t} - I_{i,t-1},, \quad \forall\ i,\ t \tag{27}$$

$$y_{i,t} + z_{i,t} \leq 1, \quad \forall\ i,\ t \tag{28}$$

Les contraintes des coûts de démarrage et d'arrêt sont données comme suit:

$$SU_{i,t} \geq CU_{i,t}(I_{i,t} - I_{i,t-1}), \quad \forall\ i,\ t \tag{29}$$

$$SU_{i,t} \geq 0, \quad \forall\ i,\ t \tag{30}$$

$$SD_{i,t} \geq CD_{i,t}(I_{i,t-1} - I_{i,t}), \quad \forall\ i,\ t \tag{31}$$

$$SD_{i,t} \geq 0, \quad \forall\ i,\ t \tag{32}$$

## Les Contraintes Thermiques

$$T_{j,t+1}^s = A_j^{\mathsf{d}} T_{j,t}^s + B_j^{\mathsf{d}} U_{j,t}^s, \quad \forall j,\ t,\ s \tag{33}$$

$$T_{j,t}^{s,\mathsf{in}} = C_j^{\mathsf{d}} T_{j,t}^s, \quad \forall j,\ t,\ s \tag{34}$$

$$T_{j,t}^d - \delta_{j,t} \leq T_{j,t}^{s,\mathsf{in}} \leq T_{j,t}^d + \delta_{j,t}, \quad \forall j,\ t,\ s \tag{35}$$

$$0 \leq P_{j,t}^{s,\mathsf{hvac}} \leq P_j^{\mathsf{hvac,max}}, \quad \forall j,\ t,\ s \tag{36}$$

## Les Contraintes de la Batterie

Les contraintes (37)–(40) capture les limites de la puissance de chargement et de déchargement ainsi que le niveau d'énergie stockée dans une batterie $k$. Ici, le niveau de stockage de la batterie à la fin de l'horizon de la planification est égal à son niveau d'énergie initial. Les contraintes (41)-(42) sont imposées pour assurer que la batterie ne peut pas être chargée et déchargée simultanément dans n'importe quel tranche de temps. Le modèle dynamique de l'énergie de la batterie $k$ est capturée dans (43).

$$0 \leq P_{k,t}^{s,c} \leq b_{k,t}^{s,c} P_k^{c,\text{max}}, \quad \forall k, \ t, \ s \tag{37}$$

$$0 \leq P_{k,t}^{s,d} \leq b_{k,t}^{s,d} P_k^{d,\text{max}}, \quad \forall k, \ t, \ s \tag{38}$$

$$E_k^{\text{min}} \leq E_{k,t}^s \leq E_k^{\text{max}}, \quad \forall k, \ t, \ s \tag{39}$$

$$E_{k,NH}^s = E_{k,1}^s, \quad \forall k, \ t, \ s \tag{40}$$

$$b_{k,t}^{s,c} + b_{k,t}^{s,d} = 1, \quad \forall k, \ t, \ s \tag{41}$$

$$b_{k,t}^{s,c}, \ b_{k,t}^{s,d} \in \{0, \ 1\}, \quad \forall k, \ t, \ s \tag{42}$$

$$E_{k,t+1}^s = E_{k,t}^s + (\eta_k^c P_{k,t}^{s,c} \Delta T - \frac{P_{k,t}^{s,d} \Delta T}{\eta_k^d}), \quad \forall k, \ t, \ s. \tag{43}$$

## Déversement Involontaire de Charge

$$0 \leq LS_t^s \leq LS_t^{\text{max}}, \quad \forall \, t, \ s \tag{44}$$

$$\frac{\sum_{s=1}^{NS} \rho_s LS_t^s}{\sum_{s=1}^{NS} \rho_s L_t^s} \leq LOL_t^{\text{max}}, \quad \forall t. \tag{45}$$

## Déversement de L'énergie Renouvelable

La puissance de sortie disponible de l'unité solaire peut être calculée sur la base du rayonnement solaire et de la température ambiante. La puissance de sortie disponible d'une unité éolienne peut être calculée sur la base de la vitesse du vent et de la courbe de puissance du vent. Dans chaque intervalle de temps, la quantité du vent / déversement d'énergie solaire doit évidemment être plus petit que la quantité du vent/puissance de sortie d'énergie solaire. Par conséquent, nous devons avoir.

$$0 \leq P_{w,t}^{s,\text{ws}} \leq P_{w,t}^s, \quad \forall \, w, \ t, \ s. \tag{46}$$

$$0 \leq P_{p,t}^{s,\text{pvs}} \leq P_{p,t}^s, \quad \forall \, p, \ t, \ s. \tag{47}$$

La programmation d'énergie et le problème d'appel d'offres pour l'optimisation conjointe des systèmes de CVC et des ressources distribuées dans le MR décrit dans la section précédente ,est un programme linéaire en nombres entiers mixtes (PPIM), qui peut être résolu efficacement en utilisant des solutionneurs commerciaux disponibles tels que CPLEX.

Des résultats numériques extensifs sont présentés pour illustrer les grands avantages de notre conception dans la réduction du déversement d'énergie renouvelable, l'atténuation de la haute pénalité due à un déséquilibre de l'énergie, l'augmentation du bénéfice attendu du MR en exploitant la dynamique thermique du bâtiment. Sur la base de ces études numériques, les conclusions suivantes sont dans l'ordre.

1. La coordination de la charge de CVC (ou dans les charges générales flexibles) et SER à travers un cadre unifié de gestion d'énergie est important. Intégration de la planification souple de la charge du système de CVC dans les décisions de gestion de l'énergie de l'agrégateur MR peut augmenter considérablement son bénéfice attendu et réduire la quantité de déversement de l'énergie renouvelable, qui permet également d'éviter la charge élevée du déséquilibre énergétique causé par l'écart de la soumission

2. Les avantages du dispositif coordonné proposé dépendent de la flexibilité offerte par le système de CVC. Plus précisément, le bénéfice attendu de l'agrégateur MR augmente et la quantité de déversement d'énergie renouvelable diminue à mesure que les maximales admissibles d'écart de température, la puissance nominale du CVC systèmes augmentent et le coût de l'écart de température diminue. Cependant, le gain de performance est saturé à certaines valeurs de l'écart maximal admissible de la température, du coût d'écart de température et de la puissance nominale du CVC.

3. Le coût de la déviation de la soumission et le coût de déversement d'énergie renouvelable ont des impacts significatifs sur la solution optimale. En particulier, le bénéfice prévu du MR diminue et la quantité de déversement d'énergie renouvelable augmente comme le coût de la déviation de la soumission augmente et/ou le coût de déversement d'énergie renouvelable augmente aussi.

4. Le stockage de la batterie peut aider l'agrégateur MR à réduire la quantité de déversement d'énergie renouvelable et à augmenter le bénéfice attendu. Le bénéfice anticipé d'agrégateur MR augmente et le montant de déversement des énergies renouvelables diminue comme la limite d'échange de puissance maximale entre l'agrégateur MR et réseaux principaux augmente. Cependant, elle est saturée à mesure cette limite maximale de puissance devient suffisamment grande. Enfin, le niveau d'incertitude a des impacts significatifs sur la solution optimale.

# Abstract

More active participation of the demand side and efficient integration of distributed energy resources (DERs) such as electric vehicles (EVs), energy storage (ES), and renewable energy sources (RESs) into the existing power systems are important design objectives of the future smart grid. In general, effective demand side management (DSM) would benefit both system operators (e.g., peak demand reduction) and electricity customers (e.g., cost saving). For building and home energy scheduling design, heating, ventilation, and air-conditioning (HVAC) systems play a very important role since HVAC power consumption is very significant and the HVAC load can be scheduled flexibly while still maintaining user comfort requirements. This thesis focuses on energy scheduling design for two different application scenarios where HVAC and various DERs are considered to optimize the benefits electric users.

The first part of the thesis studies the joint scheduling optimization of EVs and HVACs, which aims to minimize the total electricity cost considering user comfort requirements. The proposed design exploits EVs as a dynamic storage facility where the energy stored in each EV can be used to charge other EVs (EV2EV) or to supply to HVAC systems (EV2HVAC) during the high-priced periods. Various system and design parameters such as user temperature comfort preference, household occupancy and EV travel patterns as well as detailed modeling of building thermal dynamics are captured in the proposed model. Under our design, optimal power consumption profiles of HVACs and optimal charging/discharging profiles of EVs can be obtained by solving a simple linear programming (LP) problem. Numerical studies show that the HVAC systems tend to consume more energy during off-peak hours to precool (preheat) buildings in summer (winter) and consume less energy during the on-peak periods while still maintaining the indoor temperature within the predefined comfort range thanks to building thermal inertia. Furthermore, our design enables a subset of EVs to be discharged to supply electricity to the HVAC systems and other EVs during on-peak hours.

This is confirmed to result in significant cost saving, allow more flexibility in setting the tradeoff between cost and user comfort, and reduce energy demand during on-peak hours.

The second part of the thesis investigates an optimal power scheduling and bidding problem for a community-scale microgrid (MG) under the day-ahead (DA) pricing. The considered MG consists of RESs (e.g., wind turbines, solar panels), conventional generating units (e.g., fuel cells, microturbines), a number of buildings with their associated loads, and an optional battery storage facility. The proposed optimization framework aims to balance between maximizing the expected benefit of the MG and minimizing the MG operation cost considering user thermal comfort requirements and other system constraints. The underlying problem is formulated as a two-stage stochastic program where first-stage decisions include commitment statues of all conventional units and hourly bid quantities that the MG aggregator submits to the DA market while the second-stage decisions comprise power dispatch decisions, actual power exchange between the MG and the main grid, battery charging/discharging decisions, and amount of involuntary load curtailment and renewable energy curtailment. The thermal dynamic characteristics of buildings is exploited to compensate for the variability of renewable energy generation. Numerical results show that integrating flexible HVAC load scheduling into energy management framework of the MG aggregator can indeed increase significantly the MG profit and reduce the amount of renewable energy curtailment, which also helps mitigate the high energy imbalance charge caused by bid deviation.

# Acknowledgements

I would like to express my deepest gratitude to my supervisor, Prof. Long Le, for supporting and guiding me during my studies at INRS-EMT, University of Quebec. I always find him an in-depth source of knowledge and I am tremendously inspired by his way of doing research. Prof. Long Le has always helped me to achieve the best in my M.Sc. studies. Without his constant support, encouragement and availability, this thesis could not have been finished.

I also would like to thank all my labmates in the Networks and Cyber Physical Systems Lab at INRS-EMT, University of Quebec, for the discussions and encouragement. They have made my study life in Montreal pleasurable.

I am also very honored to have Prof. Martin Maier, from INRS-EMT, University of Quebec, and Prof. François Bouffard, from McGill University, as members of the thesis examination committee and I want to thank them for their constructive comments and suggestions on my research.

I am grateful for the financial support from INRS-EMT, University of Quebec. This financial source greatly helps me keep working on my research and finish this thesis. Finally, I would like to thank my parents for their unconditional love and support. To them, I dedicate this thesis.

# Contents

# List of Figures

# List of Tables

# Glossary

**Abbreviations**

| | |
|---|---|
| $DA$ | Day ahead |
| $DG$ | Conventional generating unit |
| $DISCO$ | Distribution company |
| $DSM$ | Demand side management |
| $ES$ | Battery energy storage unit |
| $EV$ | Electric vehicle |
| $FC$ | Fuel cell |
| $FERC$ | Federal Energy Regulatory Commission |
| $GENCO$ | Generation company |
| $HVAC$ | Heating, ventilation, and air conditioning |
| $LMP$ | Locational marginal price |
| $LP$ | Linear programming |
| $LSE$ | Load serving entity |
| $LSF$ | Load scaling factor |
| $MG$ | Microgrid |
| $MILP$ | Mixed integer linear program |

$MT$        Microturbine

$PV$        Solar power unit

$RES$       Renewable energy sources

$RT$        Real time

$SOC$      State of charge

$USF$      Uncertainty scaling factor

$WPF$      Wind power unit

**Indices**

$i, k$         Index of DG/ES

$j$           Index of house (building)

$m$         Index of segment of piecewise linear cost function of DG $i$

$s$           Index of scenario

$t$           Index of time slot

$w, p$        Index of WP/PV

**Parameters and Constants**

$\Delta T$        Duration of time slot (h)

$\delta_{j,t}$        Maximum allowable temperature deviation of house (building) $j$ at time $t$ (°C)

$\eta_j$         Coefficient of performance (COP) of HVAC system in house (building) $j$

$\eta_p$         Conversion coefficient of PV $p$ (%)

$\eta_{j,e}^c, \eta_{j,e}^d$    Charging/discharging efficiency of EV $e$ of house $j$

$\eta_k^c, \eta_k^d$     Charging/discharging efficiency of ES $k$

$\Lambda_{j,e,l}$       Travel time of EV $e$ of house $j$ during trip $l$

$\Phi_t$        Solar irradiance at time slot $t$ (kW/m$^2$)

$\pi_{j,t}$        Cost of temperature deviation in house (building) $j$ at time slot $t$ ($/°C)

$\psi_t$        Cost of bid deviation at time slot $t$ ($/kWh)

$\rho_s$        Probability of scenario $s$

$\sigma_j$        Dummy variable, "-1" for AC cooling, "1" for heating

$a_{j,t}$        Occupancy state of house $j$ at time slot $t$, "1" for occupied, "0" otherwise

$b_{j,e,t}$        Availability of EV $e$ of house $j$ at time slot $t$, "1" for parking at home, "0" otherwise

$C$        Thermal capacitance (kWh/°C)

$C_k^{\text{deg}}$        Operating (degradation) cost of ES $k$ ($/kWh)

$CD_{i,t}, CU_{i,t}$        Shutdown/startup offer cost of DG $i$ ($)

$d_{j,e,l}$        Travel distance of EV $e$ of house $j$ during trip $l$ (km)

$DR_i, UR_i$        Ramping-down/up rate limit of DG $i$ (kW)

$DT_i, UT_i$        Minimum down/up time of DG $i$ (h)

$E_k^{\text{cap}}$        Capacity of ES $k$ (kWh)

$E_k^{\min}, E_k^{\max}$        Minimum/maximum energy stored in ES $k$ (kWh)

$e_t$        Electricity price at time slot $t$ ($/kWh)

$E_{j,e}^{\text{cap}}$        Battery capacity of EV $e$ of house $j$ (kWh)

$LOL_t^{\max}$        Maximum loss of load percentage at time slot $t$

$LS_t^{\max}$        Maximum load shedding (kW)

$m_{j,e}$        Travel efficiency of EV $e$ of house $j$ (kWh/km)

$N_i$        Number of segments of piecewise linear cost function of DG $i$

$NB$        Number of households (buildings)

$NE_j$      Number of EVs in house $j$

$NG, NK$      Number of DGs/ESs

$NH$      Number of time slots

$NS$      Number of scenarios

$NW, NP$      Number of WPs/PVs

$P_i^{\mathsf{min}}, P_i^{\mathsf{max}}$      Minimum/maximum power generation of DG $i$ (kW)

$P_j^{\mathsf{hvac,max}}$      Power rating of HVAC system in house $j$ (kW)

$P_j^{\mathsf{hvac,max}}$      Rated power of HVAC system $j$ (kW)

$P_w^{\mathsf{r}}$      Rated power of WP $w$ (kW)

$P_{j,e}^{\mathsf{ev,c,max}}$      Maximum charging power of EV $e$ of house $j$ (kW)

$P_{j,e}^{\mathsf{ev,d,max}}$      Maximum discharging power of EV $e$ of house $j$ (kW)

$p_j^{\mathsf{sr}}$      The fraction of solar radiation entering the inner walls and floor of house $j$

$P_k^{\mathsf{c,max}}, P_k^{\mathsf{d,max}}$      Maximum charging/discharging power of ES $k$ (kW)

$R$      Thermal resistance between two heat exchange media ($^o$C/kW)

$S_p$      Array area of PV $p$ (m$^2$)

$SOC_{j,e}^{\mathsf{max}}$      Maximum allowable SOC of EV $e$ of house $j$

$SOC_{j,e}^{\mathsf{min}}$      Minimum allowable SOC of EV $e$ of house $j$

$T_t^{\mathsf{a}}$      Ambient temperature at time slot $t$ ($^o$C)

$t_{j,e,l}^{(1)}$      Time slot index when EV $e$ of house $j$ leaves home for trip $l$

$t_{j,e,l}^{(2)}$      Time slot index when EV $e$ of house $j$ returns home for trip $l$

$T_{j,t}^{\mathsf{d}}$      Desired indoor temperature of building $j$ at time slot $t$ ($^o$C)

$V^{\mathsf{RES}}$      Cost of renewable energy curtailment ($/kWh)

$V_t^{\mathsf{W}}, V_t^{\mathsf{PV}}, V_t^{\mathsf{LL}}$      Cost of wind/solar/load curtailment ($/kWh)

$v_w^r, v_w^{ci}, v_w^{co}$      Rated/cut-in/cut-out wind speed (m/s)

$w_{j,t}$      Weighting coefficient between electricity cost and discomfort cost for house $j$ at time slot $t$ ($/°C)

## Stochastic Parameters

$\Phi_t^s$      Solar irradiance at time slot $t$ in scenario $s$ (kW/m$^2$)

$e_t^{s,\text{DA}}, e_t^{s,\text{RT}}$      DA/RT electricity prices at time slot $t$ in scenario $s$ ($/kWh)

$L_t^s$      Total non-HVAC load at time slot $t$ in scenario $s$ (kW)

$P_{p,t}^s$      Available output power of PV $p$ at time slot $t$ in scenario $s$ (kW)

$P_{w,t}^s$      Available output power of WP $w$ at time slot $t$ in scenario $s$ (kW)

$T_t^{s,\text{a}}$      Outdoor temperature at time slot $t$ in scenario $s$ (°C)

$v_t^s$      Wind speed at time $t$ in scenario $s$ (m/s)

## Variables

$P_t^{\text{grid}}$      Imported power from the grid at time slot $t$ (kW)

$P_{j,e,t}^{\text{ev},c}$      Charging power of EV $e$ of house $j$ at time slot $t$ (kW)

$P_{j,e,t}^{\text{ev},d}$      Discharging power of EV $e$ of house $j$ at time slot $t$ (kW)

$P_{j,t}^{\text{hvac,out}}$      Output power of HVAC system in building $j$ at time slot $t$ (kW)

$P_{j,t}^{\text{hvac}}$      Power supplied to HVAC system in building $j$ at time slot $t$ (kW)

$SOC_{j,e,t}$      State of charge of EV $e$ of house $j$ at time slot $t$

$T_{j,t}^{\text{e}}$      The temperature of the building envelope (°C)

$T_{j,t}^{\text{in}}$      Indoor temperature of building $j$ at time slot $t$ (°C)

$T_{j,t}^{\text{m}}$      The temperature of the thermal accumulating layer in the inner walls and floor (°C)

## First-stage Variables

$I_{i,t}$          Commitment status of DG $i$ at time slot $t$ {0, 1}

$P_t$          Scheduled DA bid quantity at time slot $t$ (kW)

$SD_{i,t}, SU_{i,t}$          Shutdown/startup cost of DG $i$ at time slot $t$ ($)

$y_{i,t}, z_{i,t}$          Startup and shutdown indicators {0, 1}

**Second-stage Variables**

$b_{k,t}^{s,c}, b_{k,t}^{s,d}$          Binary variable, "1" if ES $k$ is charged/discharged at time slot $t$ in scenario $s$, "0" otherwise

$C_i(.)$          Production cost of DG $i$ ($)

$E_{k,t}^s$          Energy stored in ES $k$ at time slot $t$ in scenario $s$ (kWh)

$LS_t^s$          Realized load shedding at time slot $t$ in scenario $s$ (kW)

$P_t^s$          Actual power delivery at time slot $t$ in scenario $s$ (kW)

$P_{i,t}^s$          Power generation of DG $i$ at time slot $t$ in scenario $s$ (kW)

$P_{i,t}^s(m)$          Power generation of DG $i$ from the $m$-th segment at time slot $t$ in scenario $s$ (kW)

$P_{j,t}^{s,\text{hvac,o}}$          Output power of HVAC system $j$ at time slot $t$ in scenario $s$ (kW)

$P_{j,t}^{s,\text{hvac}}$          Input power supplied to HVAC system $j$ at time slot $t$ in scenario $s$ (kW)

$P_{k,t}^{s,c}, P_{k,t}^{s,d}$          Charging/discharging power of ES $k$ at time slot $t$ in scenario $s$ (kW)

$P_{w,t}^{s,\text{ws}}, P_{p,t}^{s,\text{pvs}}$          Wind/solar power curtailment at time slot $t$ in scenario $s$ (kW)

$T_{j,t}^{s,\text{in}}$          Indoor temperature of building $j$ at time slot $t$ in scenario $s$ (ºC)

# Chapter 1

# Introduction

## 1.1 Motivation

Dealing with the ever-increasing demand for electricity has always been a significant challenge for electric utilities. An obvious solution to this problem is to expand the grid capacity to meet increasing demand. However, this strategy has limitations since the capital investment to build new power plants, transmission and distribution facilities can be very costly. An alternative solution is to utilize the existing grid infrastructure in a more efficient way. In fact, various new concepts in the smart grid domain have emerged to tackle this challenging problem. In general, smart grid technologies could help improve the grid reliability, reduce peak power demand and therefore allow deferral of major and costly upgrades. In addition, more advanced information and communications infrastructure integrated into the electricity network, which facilitates real-time interactions among different entities in the system in the smart power grid [1, 2]. Therefore, utilities or energy service providers can directly communicate with their customers via smart meters to exchange information such as electricity price and power consumption which benefits both the supply and demand sides. In particular, based on customer power usage data, utilities can estimate the system load more accurately based on which they can schedule their operation more efficiently. Also, with the electricity price information, customers have the opportunity to reduce their bills by shifting their power consumption to low price hours. Furthermore, modern communications infrastructure and sensing and metering technologies enable the system operator to obtain real-time information about the grid conditions (e.g., voltage and frequency information, fault detection) to enhance the system efficiency and reliability [2]. Finally, smart grid technologies would ease the deployment of advanced energy management techniques such as optimization of energy usage and network resources as well as active participation of demand side into the operation of the power system.

Among grid operations and functions, balancing electricity supply and demand in real time is one of the most critical tasks to ensure the grid stability and power quality. Traditionally, the electricity system is built in such a way that some spare capacity is available to accommodate peak power demand. As a result, some important grid infrastructure (e.g., power plants) can be underutilized for most of the time except for certain peak usage times which can occur rarely. Thus, it is very desirable to lower peak demand to reduce the burden on upfront cost of ungrading grid facilities. Furthermore, as a result of the market clearing procedure, electricity prices and power demands are strongly correlated where electricity price is typically high during high load periods and vice versa. Consequently, peaker power plants are brought online to cope with daily peak demand. This not only rises the wholesale electricity price due to high marginal cost of these power plants but also harms the environment by increasing carbon emissions. Efficient peak demand management is, therefore, crucial in grid expansion planning and maintaining reasonable electricity prices.

Instead of increasing power generation capacity, demand side management (DSM) has been considered as an alternative solution for managing power demand. DSM aims to improve energy efficiency (e.g., using more efficient appliances) and enable demand response (e.g., load curtailment, load shifting) [3]. According to the U.S. Federal Energy Regulatory Commission (FERC), demand response (DR) is described as *"changes in electric usage by end-use customers from their normal consumption patterns in response to changes in the price of electricity over time, or to incentive payments designed to induce lower electricity use at times of high wholesale market prices or when system reliability is jeopardized"* [4]. Many DSM programs have been offered by electric utilities to encourage energy users to adjust the consumption level and usage pattern. The ultimate goal of DR programs via either *incentive-based DR* or *price-based DR* program is to redistribute electricity load and lower peak demand. Furthermore, the release of Order 745 issued recently by FERC [5] allows customers to bid their DR capacity directly in the wholesale energy market and DR bids are treated in the same manner as other market participants. Interested readers can refer [3, 4, 5, 6, 7, 8] for further details about various DSM programs. In summary, DSM not only helps utilities and system operators defer the need of upgrading their facilities and enhance customer services but also provide customers opportunity to lower their electric bills.

Among household and building loads, HVAC is the primary contributor to peak load in summer and winter months [9]. In general, electricity demand changes over time and follows daily, weekly, and seasonal patterns [10]. For example, in a hot summer day or a cold winter day, the peak demand would be high since homes and buildings would utilize the HVAC system at the maximum capacity. Smart energy scheduling for HVAC systems is an important research topic. In terms of HVAC scheduling constraint, users typically desire to maintain the indoor temperature at a certain preferred setpoint, which depends on their preferences and occupancy status [11]. In addition, users can tolerate a small deviation of the indoor temperature from the desired setpoint. However, users would feel more comfortable when the indoor temperature is closer to the preferred setpoint. Here, the HVAC power consumption can be scheduled intelligently to achieve electricity cost saving without violating users' comfort requirements. In particular, the HVAC system can consume more power during low-priced hours to precool (preheat) buildings in the summer (winter) and it can reduce the power consumption during high-priced hours while still maintaining indoor temperature in the comfort zone thanks to the building thermal inertia.

The main reasons that HVAC power scheduling can play an important role in DSM can be summarized as follows. First, HVAC load contributes a significant portion of total energy consumption in residential and commercial buildings. Consequently, it accounts for a major part in customers' electric bills. Second, the aggregated power consumption of HVAC systems is the main cause of summer and winter peak loads. Finally, despite being one of the most enery-hungry appliances, the HVAC system offers a great flexibility in controlling its power consumption while respecting users' comfort requirements. In the following, we provide a literature survey for the home energy scheduling topic and discuss the motivation for our research work in this thesis.

## 1.2   Literature Review

Demand side management in buildings, including both residential and commercial buildings, is an important research topic which has received lots of attention from the research community [12, 13, 14, 15, 16, 17, 18]. This is partly because buildings contribute for a significant fraction of overall electricity consumption, which accounted for 72% of total U.S. energy consumption in 2006 and out of that residential buildings accounted for 51% according to the U.S. Environmental Protection Agency (EPA) [19]. In [12], Shao *et al.* assessed the potential of DR as a load shaping tool to improve the distribution transformer utilization and avoid overloading such transformers. Mohsenian-Rad *et al.* [13] proposed an optimization framework that aims to minimize electric bills considering user comfort. However, the assumption on homogeneous appliances and using waiting time to represent user comfort in this paper would be too simple to represent different characteristics of home appliances and user requirements.

In a typical household, thermostatically controlled appliances (TCAs) including refrigerators, electric water heaters, and HVAC systems account for more than half of total residential energy consumption [19]. Research on optimal control for TCA loads has been a hot research topic over the last several years. References [14] and [15] proposed optimal control schemes to minimize the electricity cost for the HVAC system considering user climate comfort. Dynamic programming was employed in [16] to compare several optimal control algorithms applied to a thermostat. In [17], the authors introduced an appliance commitment algorithm that schedules electric water heater power consumption to minimize user payment. Kondoh *et al.* investigated the potential of using water heater load to provide regulation service in [18].

Electric Vehicles (EVs) are other important grid elements that have significant potential economic and environmental advantages compared to regular cars. The penetration of EVs is expected to increase drastically in the next few years, which could reach one million by 2015 in the U.S. [20]. Therefore, EV charging will have significant impacts on power distribution networks if it is not controlled appropriately [21, 22, 23]. EV travel pattern is an important factor to model potential impacts of EVs on the grid [24, 25] and to develop efficient EV charging strategies [26]. Given electricity prices and EV driving patterns, Rotering *et al.* proposed a dynamic programming based control scheme to optimize the charging for one EV [27]. In [28], Wu *et al.* considered load scheduling and dispatch problem for a fleet of EVs in both the day-ahead market and real-time energy market. In [29], an optimal charging strategy for EVs was proposed that considers voltage and power constraints. The problems of scheduling of home energy usage and EV charging are often addressed separately in the literature. In this thesis, a unified optimization model is proposed to jointly optimize the energy scheduling of EVs and HVACs, which will be presented in chapter 3.

There has been also growing interest in integrating more distributed energy resources (DERs), especially renewable energy sources (RESs), into the future smart grid by exploiting advanced communications and control technologies. In fact, RESs have various advantages over the traditional power sources since they are the green sources of power with almost-zero operating and emissions costs. However, the intermittent and volatile nature of renewable energy generation imposes a significant challenge to integrate these resources into the power system. Various methods have been proposed to tackle the intermittency and volatility of RESs. In [30, 31, 32, 33], the authors proposed to employ pumped-storage hydro units coordinated with wind sources to maximize the profits of generation companies (GENCOs), or to minimize the operating costs for power system operators. The optimal battery sizing problem was studied in [34, 35] to cope with the uncertainties in renewable energy generation. The potential of using charging/discharging capability of electric vehicles (EVs) to support renewable energy was investigated in [36, 37]. Other technologies such as compressed air storage [38], fast-response units (e.g., gas-fired units) [39] can also be used to mitigate the fluctuation of renewable energy generation. However, these solutions have some drawbacks, e.g., pumped-storage hydro systems are geographically dependent and it takes a long time to build a pumped-storage facility with a high capital cost.

In chapter 4 of the thesis, we consider the potential of using HVAC systems to cope with the challenging problem of renewable energy integration. As mentioned above, there have been several proposed optimal control schemes for HVAC systems to minimize their operation costs considering different temperature comfort criteria [11, 40, 41, 42, 43]. In the proposed solution, we not only consider electricity price variation in scheduling HVAC power consumption but we also propose to exploit the thermal dynamic characteristics of buildings (e.g., campus, residential or office buildings) to mitigate the uncertainties of renewable energy generation. The flexibility offered by building thermal storage capability [1] could make the HVAC system a great candidate in solving the renewable energy integration problem. The idea is that HVAC power consumption can be scheduled according to the renewable energy generation profile, e.g., when the renewable power is higher than expected, HVAC systems may consume more power to precool/preheat buildings and vice versa. The potential of the thermal storage capability of buildings is assessed in a MG setting where the considered MG participates in a deregulated electricity market with the objective of maximizing its expected profit (i.e., revenue minus operation cost).

Optimal bidding strategies for various entities in power markets have been extensively studied in the literature. The majority of existing works has focused on optimizing the operation of generators in the supply side to maximize the profits of GENCOs [30, 31, 32, 33, 44, 45, 46, 47, 48, 49, 50]. In addition to GENCOs, there are other important entities such as distribution companies (DISCOs), retail companies (RETAILCOs), aggregators [44] that can participate in the deregulated electricity market where these entities can be considered belonging to the demand side (e.g., buying electricity from the wholesale market to serve customers). In [51, 52, 53, 54, 55], different operation frameworks for DISCOs, RETAILERs, and larger customers in the competitive electricity market have been proposed. In [56, 57], the authors studied the optimal energy trading problem for an aggregator that controls the operation of a number of EVs.

---

[1] The thermal storage capability of each building depends on the building architecture and properties (e.g., construction materials, window area).

In our proposed solution, we consider a particular market entity which is a MG aggregator. In general, a MG can be defined as a cluster of DERs and associated loads, and it can be operated in grid-connected mode or islanded mode [58]. A MG can be of different sizes ranging from a building, a university campus, to the community scale (e.g., a village). Capacities of DERs in a MG can be relatively small to allow them to participate directly in the power market. Therefore, local electricity generation and demand in the MG can be aggregated and controlled by a MG aggregator which serves as the representative of the MG in the market. Moreover, the MG can be considered as a *"prosumer"* which not only consumes but also produces electricity [59]. Particularly, when the local generation is higher than the local demand, the MG acts as a producer selling its surplus energy to the main grid. In contrast, when the local generation is not sufficient to meet the local demand, the MG plays the role of a consumer who buys electricity from the market to serve its local demand. In this work, we assume that the MG is allowed to participate in the electricity market and the market operator treats the MG the same manner as other market entities (e.g., GENCOs and DISCOs). From the market operator's perspective, the MG can act either as a supplier or a customer depending on the direction of power flow between the MG and the main grid.

There is an important difference between MGs and GENCOs in that the supply-demand balancing constraint does not exist in the power scheduling problem for GENCOs. On the other hand, supply-demand balance is a critical requirement in the power scheduling and bidding problem for MGs. Moreover, flexible loads can be integrated in the power scheduling optimization for MGs, which can potentially reduce considerably the operating cost for MGs and compensate for the fluctuation of renewable energy generation. Optimal energy trading for MGs has been considered in several studies [58, 60, 61, 62, 63, 64, 65, 66, 67, 68, 69, 70, 71] where the typical objectives are to maximize the revenue for the MG in the power market and to minimize the MG operation cost. The work in this thesis belongs to this line of research, which, however, has several distinct modeling aspects, which will be described in details in chapter 4.

## 1.3    Research Objectives and Contributions

The main objective of the research pursued in this thesis is to develop smart energy scheduling algorithms for customers' buildings considering special operation characteristics of HVAC, building thermal dynamics, and users' comfort preference. The development of such scheduling algorithms are studied in two different application scenarios. Furthermore, detailed operation modeling of different DERs, the concept of MG, and the energy trading principles in the competitive electricity market are considered. Specifically, the contributions of this thesis are described the following.

The first part of this thesis focuses on the joint scheduling optimization of EVs and HVACs, which is presented in chapter 3. The scheduling problems of home energy usage and EV charging are often addressed separately in the literature. In this chapter, a unified optimization model is presented to jointly optimize the scheduling of EVs and HVACs [11]. In particular, EVs are utilized as a dynamic storage facility to supply energy for residential buildings during peak hours where the energy can be transferred from EVs to charge other EVs and to provide energy for HVACs. The proposed model aims to minimize the total electricity cost considering user comfort, house occupancy, EV travel patterns, thermal dynamics, EV electricity demand, and other operation constraints.

HVAC and EV loads contribute for a significant portion of the total power consumption in the residential sector. Furthermore, these appliances offer some beneficial scheduling flexibility. For other appliances, adjusting their operation schedule might lead to considerable inconvenience for users with minor gain in electricity cost saving. Hence, only HVAC and EV are considered in the proposed model, which is not only sufficient to draw insightful conclusions about the interaction of EVs and building energy management but also allows us to study detailed interactions of these two major power-hungry appliances and the impact of different design and system parameters on the optimal solution. Nevertheless, the proposed model can be extended to include other appliances such as flexible loads (e.g., water heaters, washers, dryers) and non-flexible loads (e.g., TV, lighting) [43]. The main contributions of the first part of this thesis can be summarized as follows:

- A comprehensive optimization model is proposed to optimize the EV and HVAC scheduling in a residential area. The formulation aims to achieve flexible tradeoff between minimizing total electricity cost and maintaining user comfort preference. It also accounts for the characteristics of the HVAC system, thermal dynamics, user comfort preference, battery state model, user travel patterns, and household occupancy patterns. Potential extensions of the proposed framework to capture various modeling uncertainty factors are also discussed.

- The impacts of different design and system parameters, which control the electricity cost and user comfort, on the system operation and performance as well as the economic benefits of applying the proposed control framework compared to a non-optimized control scheme for a single-house scenario are presented.

- The advantages of applying the proposed control model for the multiple-house scenario compared to the case where each household optimizes its energy consumption separately is illustrated. Specifically, we show that optimization of EV and home energy scheduling for multiple houses in a residential community can achieve significant cost saving and reduce the high power demand during peak hours.

The second part of this thesis considers the integration of HVAC energy scheduling into an energy management framework of a renewable-powered MG, which is presented in chapter 4. To best of our knowledge, there is no existing economic energy bidding framework that evaluates the potential of exploiting HVAC systems and the associated thermal load to cope with the uncertainties of RESs and maximize the profit for MGs in the electricity market. Our work fills this important gap where its main contributions can be summarized as follows:

- We propose a comprehensive model based on which we develop an optimal DA scheduling strategy for a MG in a two-settlement electricity market, which is the common practice in the U.S. [33, 44, 49]. The proposed model aims to balance between maximizing the revenue for the MG and minimizing the load and renewable curtailment as well as bid deviation while maintaining users' comfort requirements and other system constraints. The proposed model is novel in that it enables us to exploit the building thermal dynamics properties to compensate for the variability of renewable energy generation, which can significantly improve the MG profit. Specifically, HVAC load is used as a DR source, which is integrated into the optimal bidding strategy of the MG aggregator in the electricity market.

- The optimization model is formulated as a two-stage stochastic programming problem where uncertainties are captured by using the Monte Carlo simulation method. First-stage and second-stage variables are appropriately defined for efficient operations of the power systems integrating different renewable energy sources.

- We present extensive simulation results to demonstrate the advantages in coordinating the operations of the HVAC systems and renewable energy resources compared to the uncoordinated case where HVAC systems and other components of the MG optimize their power consumption/generation independently. The performance of the proposed scheme is also compared with that under the strict climate comfort requirement where no temperature deviation is allowed. Finally, the sensitivity analysis is performed to assess the impacts of different system and design parameters on the optimal solution.

## 1.4 Thesis Outline

The rest of this thesis is organized as follows. Chapter 2 covers the relevant background to this thesis including the modeling of building thermal dynamics and basic optimization techniques including linear programming, stochastic optimization, and some popular optimization toolboxes.

Chapter 3 presents the system model and mathematical formulation of the join scheduling optimization framework for EVs and HVACs. Furthermore, this chapter also discusses briefly the potential of using Model Predictive Control (MPC) to tackle system uncertainties and the integration of other types of residential loads into the proposed optimization framework. Finally, numerical results are presented to illustrate the effectiveness of the proposed design.

Chapter 4 describes a stochastic optimization framework for energy scheduling and bidding activities of microgrids (MGs) where the HVAC load is utilized to compensate for the uncertainties in the bidding decision process of the MG aggregator. Detailed modeling of various components in the proposed MG model are addressed in this chapter. The underlying optimization problem is formulated as a two-stage stochastic programming problem. The proposed framework enables the MG aggregator to make optimal bidding decisions in the DA energy market as well as optimal decisions on the real-time operation of MG components including power dispatch, load and renewable energy curtailment, and actual power consumption profiles of HVACs.

Finally, chapter 5 concludes the thesis by providing a brief summary of the accomplished work and outlining some future research directions.

# Chapter 2

# Background

This chapter presents the general background relevant to the design problems considered in this thesis. First, a detailed mathematical model of building thermal dynamics is presented, which describes the relationship between HVAC power consumption and building indoor air temperature. Then, we review basic optimization techniques including linear programming, mixed integer linear programming, stochastic optimization and optimization toolboxes used in the thesis.

## 2.1 Building Thermal Dynamics

To formulate scheduling optimization problems in the next chapters, we need a model representing the indoor air temperature dynamics and HVAC load. In fact, modeling thermal dynamic of buildings is an important research topic that has been extensively studied in the literature. Among existing modeling methods for building thermal dynamics, the grey-box approach appears to be one of the most popular ones. In this approach, we combine the physical knowledge about the building and experimental data to obtain a reasonable model for the building thermal dynamics [40, 41, 72, 73, 74, 75, 76]. Based on the energy balance and mass balance equations for the indoor air, a continuous time linear state space model, which is a set of first-order differential equations, can be constructed [42, 75, 76]. It is also called equivalent thermal parameter model (ETP) [17]. Then, experimental data is used to estimate thermal parameters of the constructed model [41, 73, 74, 75, 76].

In the following, we present a commonly used thermal dynamic model for buildings. For building $j$, the indoor temperature can be expressed as a function of the housing thermal characteristics (thermal resistance, thermal capacitance, window area), the weather condition (ambient temperature, solar radiation, wind speed, humidity), internal gains (e.g., occupants, cooking, refrigerator), and the HVAC input power. Specifically, we have

$$T_{j,\tau+1} = f(T_{j,\tau}, P_{j,\tau}^{\mathsf{hvac,out}}, \text{ other inputs and parameters}) \tag{2.1}$$

where $T_{j,\tau}$ is the indoor temperature of building $j$ at time $\tau$ and $P_{j,\tau}^{\mathsf{hvac,out}}$ is the output power of the HVAC system in building $j$ at time $\tau$. For simplicity, we make the following assumptions.

- Each building is modeled as a large room exchanging thermal energy with the ambient environment. The indoor temperature is uniformly distributed within a building, which can be considered as the equivalent indoor temperature.

- If the building has more than one AC/heater (e.g., one AC/heater for each room) then the thermal output power from these ACs/heaters are considered as one aggregate AC/heater with the output power equal to the total output power of individual ACs/heaters [40, 41]. In practice, a central HVAC system can provide all cooling/heating loads of buildings [42, 77].

- The impact of disturbances such as humidity, internal heat gains, wind speed on the building thermal dynamics is assumed to be negligible compared to the influence of the ambient temperature, the solar radiation power, and HVAC power input.

Note that these assumptions are commonly made in the literature [41, 42, 73, 74]. The thermal energy from solar irradiance through windows can be calculated as [41]

$$Q_{j,\tau}^{\mathsf{sr}} = \Phi_\tau A_j \tag{2.2}$$

where $\Phi_\tau$ is the solar irradiance at time $\tau$ and $A_j$ is the effective window area of building $j$.

Assume that all energy flux by solar radiation through windows is absorbed by the heat accumulating layer in the inner walls and the indoor air. We define $p_j^{\mathsf{sr}}$ as the fraction of solar radiation entering the inner walls and floor of building $j$ then the rest of the solar energy is absorbed by the indoor air, i.e., we have

$$
\begin{aligned}
Q_{j,\tau}^{\mathsf{sr,wall}} &= p_j^{\mathsf{sr}} Q_{j,\tau}^{\mathsf{sr}} \\
Q_{j,\tau}^{\mathsf{sr,air}} &= (1 - p_j^{\mathsf{sr}}) Q_{j,\tau}^{\mathsf{sr}}
\end{aligned} \tag{2.3}
$$

where $Q_{j,\tau}^{\mathsf{sr,wall}}$ and $Q_{j,\tau}^{\mathsf{sr,air}}$ denote these two energy fractions, respectively. In general, heat transfer occurs when there is a temperature difference between two spaces. Thermal energy is transferred from a higher temperature space toward a lower temperature space due to conduction, convection, and radiation [41]. Based on heat transfer mechanisms, we construct the energy balance equations, which consequently result in a third order linear model for thermal dynamics of building $j$ as follows [41]:

$$C_j^{\mathrm{m}}\frac{dT_j^{\mathrm{m}}}{d\tau} = \frac{1}{R_j^{\mathrm{m}}}(T_j - T_j^{\mathrm{m}}) + A_j p_j^{\mathrm{sr}}\Phi_j$$

$$C_j^{\mathrm{e}}\frac{dT_j^{\mathrm{e}}}{d\tau} = \frac{1}{R_j^{\mathrm{e}}}(T_j - T_j^{\mathrm{e}}) + \frac{1}{R_j^{\mathrm{ea}}}(T^{\mathrm{a}} - T_j^{\mathrm{e}}) \qquad (2.4)$$

$$C_j\frac{dT_j}{d\tau} = \frac{1}{R_j^{\mathrm{a}}}(T^{\mathrm{a}} - T_j) + \frac{1}{R_j^{\mathrm{m}}}(T_j^{\mathrm{m}} - T_j) + \frac{1}{R_j^{\mathrm{e}}}(T_j^{\mathrm{e}} - T_j) + A_j(1 - p_j^{\mathrm{sr}})\Phi_j + \sigma_j P_j^{\mathrm{hvac,out}}.$$

where $\sigma_j = 1$ corresponds to the winter time and $\sigma_j = -1$ for the summer time. Other parameters are defined as

- $R_j^{\mathrm{a}}$ is the resistance between room air and the ambient (°C/kW).

- $R_j^{\mathrm{m}}$ is the thermal resistance between room air and the the thermal accumulating layer in the inner walls and floor (°C/kW).

- $R_j^{\mathrm{e}}$ is the thermal resistance between room air and the the house envelope (°C/kW)

- $R_j^{\mathrm{ea}}$ is the thermal resistance between the house envelope and the the ambient (°C/kW).

- $C_j$ is total thermal capacitance of the indoor air (kWh/°C).

- $C_j^{\mathrm{m}}$ is the total thermal capacitance of the inner walls (kWh/°C).

- $C_j^{\mathrm{e}}$ is the total thermal capacitance of the house envelope (kWh/°C).

The equivalent thermal parameters $R_j^{\mathrm{a}}$, $R_j^{\mathrm{m}}$, $R_j^{\mathrm{e}}$, $R_j^{\mathrm{ea}}$, $C_j$, $C_j^{\mathrm{m}}$, $C_j^{\mathrm{e}}$, and $p_j^{\mathrm{sr}}$ are assumed to be constant, which can be estimated by using the Maximum Likelihood (ML) method based on measured data [41, 73, 74, 76]. Therefore, the thermal dynamics of $j$-th building can be rewritten in the deterministic linear state space model in continuous time as

$$\begin{aligned} \frac{dT_j}{d\tau} &= A_j T_j + B_j U_j \\ T_j^{\mathrm{r}} &= C_j T_j \end{aligned} \qquad (2.5)$$

where $T_j = [T_j^{\mathrm{in}},\ T_j^{\mathrm{m}},\ T_j^{\mathrm{e}}]'$ is the state vector and $U_j = [T^{\mathrm{a}},\ \Phi,\ \sigma_j P_j^{\mathrm{hvac,out}}]'$ is the input vector to the system. The output of interest is $T_j^{\mathrm{r}} = T_j^{\mathrm{in}}$ because we are interested in the indoor temperature, which directly impacts user climate comfort. Matrix $A_j$ represents the dynamic behavior of the system, and matrix $B_j$ captures the impact of input elements (ambient temperature, solar radiation, and HVAC power) on the system behavior. The matrices in the state space model (2.5) are given as follows:

$$A_j = \begin{bmatrix} a_{11} & a_{12} & a_{13} \\ a_{21} & a_{22} & 0 \\ a_{31} & 0 & a_{33} \end{bmatrix}, \quad B_j = \begin{bmatrix} b_{11} & b_{12} & b_{13} \\ 0 & b_{22} & 0 \\ b_{31} & 0 & 0 \end{bmatrix}, \quad C_j = \begin{bmatrix} 1 & 0 & 0 \end{bmatrix} \quad (2.6)$$

where the underlying coefficients are defined as

$$a_{11} = \frac{-1}{C_j}\left(\frac{1}{R_j^a} + \frac{1}{R_j^m} + \frac{1}{R_j^e}\right), \quad a_{12} = \frac{1}{R_j^m C_j}, \quad a_{13} = \frac{1}{R_j^e C_j}$$

$$a_{21} = \frac{1}{R_j^m C_j^m}, \quad a_{22} = -\frac{1}{R_j^m C_j^m}, \quad a_{31} = \frac{1}{R_j^e C_j^e}, \quad a_{33} = -\frac{1}{C_j^e}\left(\frac{1}{R_j^{ea}} + \frac{1}{R_j^e}\right)$$

$$b_{11} = \frac{1}{R_j^a C_j}, \quad b_{12} = \frac{A_j(1-p_j^{sr})}{C_j}, \quad b_{13} = \frac{1}{C_j}, \quad b_{22} = \frac{A_j p_j^{sr}}{C_j^m}, \quad b_{31} = \frac{1}{R_j^{ea} C_j^e} \quad (2.7)$$

The state space model in continuous time (2.5) can be transformed into the equivalent discrete time model by using Euler discretization (i.e., zero-order hold) with a sampling time of $T_s$ [41, 76] as

$$\begin{aligned} T_{j,\tau+1} &= A_j^d T_{j,\tau} + B_j^d U_{j,\tau} \\ T_{j,\tau}^r &= C_j^d T_{j,\tau} \end{aligned} \quad (2.8)$$

where $T_{j,\tau} = [T_{j,\tau}^{in}, \ T_{j,\tau}^m, \ T_{j,\tau}^e]'$ and $U_{j,\tau} = [T_\tau^a, \ \Phi_\tau, \ \sigma_j \eta_j P_{j,\tau}^{hvac}]'$

$$A_j^d = \exp(A T_s)$$

$$B_j^d = \int_0^{T_s} \exp(A\theta) d\theta B$$

$$C_j^d = C_j$$

This discrete time thermal dynamic model will be used in the following chapters to formulate energy scheduling optimization problems.

## 2.2 Mathematical Optimization

### 2.2.1 Basic Concepts

The standard form of an optimization problem can be expressed as follows [78]:

$$\begin{aligned}
\underset{x}{\text{minimize}} \quad & f_0(x) \\
\text{subject to} \quad & f_i(x) \leq 0, \ i = 1, \ldots, m \\
& h_i(x) = 0, \ i = 1, \ldots, p.
\end{aligned} \tag{2.9}$$

where $x \in \mathbb{R}^n$ is the optimization variable, $f_0 : \mathbb{R}^n \mapsto \mathbb{R}$ is the objective or cost function, $f_i : \mathbb{R}^n \mapsto \mathbb{R}, i = 1, \ldots, m$ are inequality constraint functions, $h_i : \mathbb{R}^n \mapsto \mathbb{R}, i = 1, \ldots, p$ are equality constraint functions. The set of $x$ that satisfies all constraints of problem (2.9) is called the feasible set. If the feasible set is empty, the problem is infeasible. The optimal value $p^*$ of the problem is defined as follows:

$$p^* = \inf\{f_0(x) | h_i(x) = 0, \ i = 1, \ldots, p, f_i(x) \leq 0, \ i = 1, \ldots, m\}. \tag{2.10}$$

Note that $p^* = \infty$ if the problem is infeasible and $p^* = -\infty$ if the problem is unbounded below.

### 2.2.2 Linear Program and Mixed Integer Linear Program

A linear program (or a linear programming problem) is a special case of the general optimization problem (2.9) where the objective function and all constraint functions are linear ones. In general, a linear program (LP) has the following form

$$\begin{aligned}
\underset{x}{\text{minimize}} \quad & \mathbf{c}'x + e \\
\text{subject to} \quad & \mathbf{A}x \leq \mathbf{b} \\
& \mathbf{C}x = \mathbf{d}
\end{aligned} \tag{2.11}$$

where

- $x \in \mathbb{R}^n$, $\mathbf{c} \in \mathbb{R}^n$, $\mathbf{b} \in \mathbb{R}^m$, $\mathbf{d} \in \mathbb{R}^p$.

- Matrices $\mathbf{A}$ and $\mathbf{C}$ have appropriate sizes.

If some or all variables $x_i$ (i.e., some or all elements in vector $x = \{x_1, x_2, ..., x_n\}$) are constrained to be integers, we have a mixed integer linear program problem (MILP).

Note that a function involving absolute operations is essentially a nonlinear function. However, if the objective function involves the absolute values of some variables, the problem can still be easily transformed to an equivalent linear programming form by introducing some auxiliary variables. For example, let us consider the following optimization problem

$$\underset{x}{\text{minimize}} \quad |\mathbf{c}'x + d| \qquad (2.12)$$
$$\text{subject to} \quad \mathbf{A}x \leq \mathbf{b}$$

where $x \in \mathbb{R}^n$. Obviously, the problem (2.12) has a nonlinear objective function. However, the problem (2.12) can be transformed into the following LP

$$\underset{t}{\text{minimize}} \quad t \qquad (2.13)$$
$$\text{subject to} \quad \mathbf{A}x \leq \mathbf{b}$$
$$\mathbf{c}'x + d \leq t$$
$$-\mathbf{c}'x - d \leq t.$$

## 2.2.3 Stochastic Programming

Stochastic programming deals with optimization problems that involve uncertainty [31, 79, 80, 81]. One of the most popular stochastic programming problems are two-stage stochastic programs where the decision variables include a set of first-stage decisions and another set of second-stage decisions. The decisions taken at the first stage, which are normally called *"here-and-now decisions"*, must be made before uncertainties are disclosed considering possible realizations of uncertain parameters at the second stage. In other words, the first-stage decisions should be based on data available at the time the decisions are made and should not depend on future observations [80]. The recourse decisions at the second stage, which are usually called *"wait-and-see decisions"*, are made after the uncertainties are unveiled, and they depend on the first-stage decisions. In summary, the solution of a two-stage program consists of a single first-stage policy and a collection of recourse decisions defining which second-stage action should be taken in response to each random outcome [31, 80]. The general formulation of a two-stage stochastic programming problem is given as follows [80]:

$$\underset{x \in X}{\text{minimize}} \quad g(x) = f(x) + E[Q(x, \xi)] \qquad (2.14)$$

where $X$ is the feasible region of $x$. For a given $\xi$, $Q(x,\xi)$ is the optimal value of the second-stage problem

$$\underset{y}{\text{minimize}} \quad q(y,\xi) \tag{2.15}$$
$$\text{subject to} \quad \mathbf{T}(\xi)x + \mathbf{W}(\xi)y \leq \mathbf{h}(\xi)$$

where $x \in \mathbb{R}^n$ is the first-stage decision variable vector, $y \in \mathbb{R}^m$ is the second-stage decision variable vector, and $\xi$ represents the uncertain data in the optimization problem.

This thesis particularly focuses on a two-stage stochastic linear program which is given by:

$$\underset{x \in X}{\text{minimize}} \quad g(x) = \mathbf{c}'x + E[Q(x,\xi)] \tag{2.16}$$
$$\text{subject to} \quad \mathbf{A}x \leq \mathbf{b}$$
$$\mathbf{C}x = \mathbf{d}$$

where $Q(x,\xi)$ is the optimal value of the second-stage problem

$$\underset{y}{\text{minimize}} \quad \mathbf{q}'(\xi)y \tag{2.17}$$
$$\text{subject to} \quad \mathbf{T}(\xi)x + \mathbf{W}(\xi)y \leq \mathbf{h}(\xi)$$

In general, a two-stage stochastic programming problem can be solved effectively by using a scenario-based optimization model. The basic idea is to convert the original stochastic optimization problem into an equivalent deterministic one. To do so, one needs to assume that the random vector $\xi$ has a finite number of possible realizations where each realization is also called a *scenario*. Suppose that the set of scenarios representing uncertain data is $\xi_1, \xi_2, ..., \xi_N$ with corresponding probabilities $p_1, p_2, \ldots, p_N$. Then, the stochastic optimization problem (2.16) can be cast as a deterministic equivalent optimization problem as follows:

$$\underset{x \in X, y_s}{\text{minimize}} \quad g(x) = \mathbf{c}'x + \sum_{s=1}^{N} p_s q(y_s, \xi_s) \tag{2.18}$$
$$\text{subject to} \quad \mathbf{A}x \leq \mathbf{b}$$
$$\mathbf{C}x = \mathbf{d}$$
$$\mathbf{T}(\xi)x + \mathbf{W}(\xi)y_s \leq \mathbf{h}(\xi), \quad \forall s$$

To generate a set of scenarios representing the uncertainties, we have to know the distributions of uncertain parameters. A popular technique used for scenario generation is the Monte-Carlo simulation method. Each scenario is a Monte Carlo sample from the distributions of uncertainties. The number of generated scenarios is chosen sufficiently large to achieve an efficient solution. In general, the larger the number of scenarios we generate, the better the optimal solution can be achieved. However, there is a trade-off between the number of simulated scenarios and the computational burden of the scenario-based optimization method. For a large-scale problem, suitable scenario reduction techniques can be employed to reduce the number of scenarios, consequently, reduce the computational burden [82, 83]. The basic idea of scenario reduction is to eliminate scenarios with very low probability and to aggregate scenarios of close distances based on certain probability metric. Further details on scenario generation approach and scenario reduction algorithms can be found in [81, 82, 83, 84].

### 2.2.4 Optimization Toolboxes

In the first part of this thesis, the joint optimization problem of EVs and HVACs is formulated as a large-scale LP problem which can be solved effectively by using CVX software [85]. In the second part of this thesis, the power bidding and scheduling optimization problem for MG is formulated using two-stage stochastic programming, which is cast as a large-scale MILP problem. We used CPLEX solver [86] to solve the problem. Scenario generation is developed in GAMS environment [87]. Furthermore, the package GAMS/SCENRED is utilized to implement the scenario reduction task.

## 2.3 Summary

This chapter discussed several relevant technical background on building thermal dynamic modeling and basic concepts of mathematical optimization. First, the discrete linear steady-state model of building thermal dynamics was introduced, which will be used later to formulate scheduling optimization problems in this thesis. Then, some basic concepts of optimization were presented. Specifically, we briefly discussed two specific classes of optimization, which are linear programming and stochastic optimization with the specific focus on two-stage stochastic programming. Finally, some relevant optimization toolboxes were also described.

# Chapter 3

# Joint Energy Scheduling Design of Electric Vehicle and HVAC

In this chapter, the energy scheduling coordination and optimization problem for HVACs and EVs is considered. The chapter begins with a description of the system model. Then, a rigorous mathematical formulation of the underlying optimization problem is introduced. Extensions of the proposed model to consider modeling uncertainties and other residential loads is also briefly discussed. Finally, case studies and numerical results are provided to show the effectiveness of the proposed model.

## 3.1 System Model

We consider the interaction among EVs and HVAC systems in a residential community. As we can see later in this Chapter, the underlying optimization model is formulated as a simple linear program which has low-complexity [88]. Therefore, the proposed optimization framework can be applied to different system sizes ranging from several households to thousands of households in the community. We assume that there is an aggregator that collects all required information from EVs and HVAC systems from all households to make control decisions. At the higher level, several aggregators can be connected to a central aggregator that coordinates the overall operations and participates in the wholesale day-ahead electricity market. The day-ahead market clearing price is assumed to be available to any aggregators before the operating day. The system model under consideration is illustrated in Fig. 3.1.

We consider a time slotted model where there are $NH$ time slots in the optimization period (e.g., 24 hours for one-day optimization period) and energy scheduling decisions are made for each time slot. The thermal dynamics and energy scheduling model that we consider in this thesis are in discrete-time, which is commonly assumed in the literature. The detailed thermal dynamics model was studied in Section 2.1. Based on the pricing information and electricity demand, each aggregator decides how much energy it should import from the grid at each time slot and how to allocate and schedule the energy usage and to exchange energy among its components including HVAC and EVs. We assume that EVs can only be charged or discharged when they are parked at home (i.e., each household is equipped with the charging facility).

The thermal inertia characteristics of buildings provides a great opportunity for DSM since the building mass can be considered as a thermal storage facility. In particular, we can schedule the power consumption of HVAC systems flexibly while respecting users climate comfort because the indoor temperature changes quite slowly. By cooperating the energy scheduling of HVAC systems and EVs, it is expected that larger cost saving can be achieved compared to the case where we schedule these loads separately. Specifically, during high price hours, the energy discharging from an EV could be used to supply for other EVs and HVAC systems or can be sold back to the main grid. In the case we do not allow to sell EV discharging power back to the grid, the discharging power from EVs is assumed to only flow within the community network. In this thesis, the term V2G (Vehicle-to-grid) refers to the case where selling back electricity to the main grid is allowed.[1]

## 3.2 Problem Formulation

We present the joint EV charging and home energy management problem for cost minimization in this section. First, the total electricity power imported from the grid at time slot $t$ can be written as

$$P_t^{\text{grid}} = \sum_{j=1}^{NB} P_{j,t}^{\text{hvac}} + \sum_{j=1}^{NB} \sum_{e=1}^{E_k} \left[ P_{j,e,t}^{\text{ev,c}} - P_{j,e,t}^{\text{ev,d}} \right], \forall t \qquad (3.1)$$

which is equal to EV charging power plus power usage of the HVAC system minus the EV discharging power summed over all households. The aggregator aims to minimize the total electricity cost and user discomfort during a scheduling horizon. We assume that there are no electricity losses in the transmission lines among EVs and HVAC systems. This assumption is reasonable since the transmission lines for energy exchange in a community are relatively short. Moreover, EV charging is assumed to be continuously controllable in our control model. The objective and constraints of the underlying optimization problem are described in the following.

---

[1]In general, V2G can refer to the case where EVs are allowed to discharge energy regardless of whether the premises become a net supplier to the main grid.

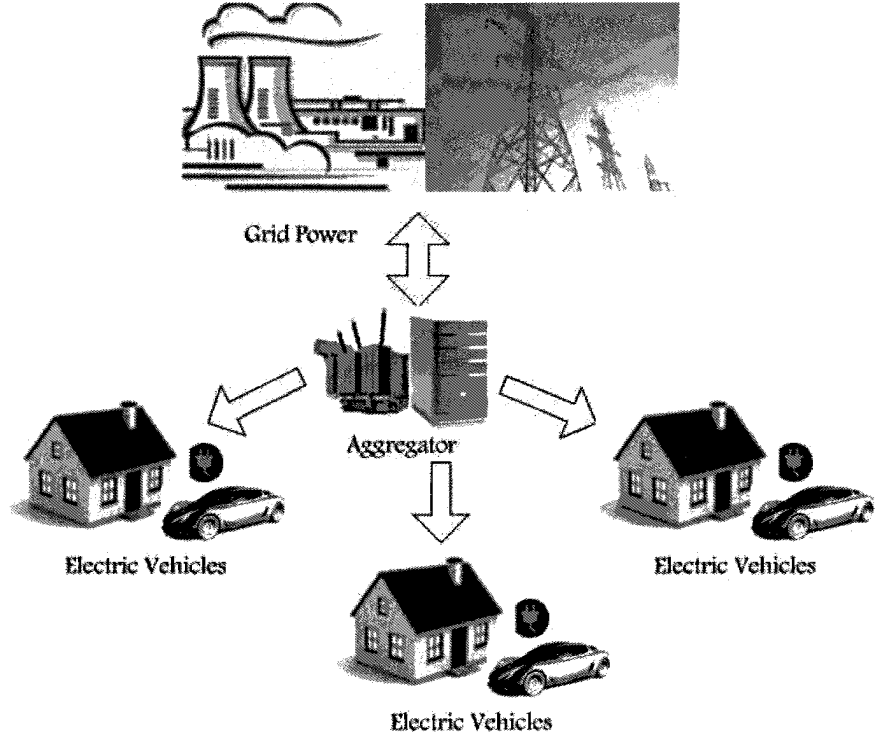

**Figure 3.1:** System model

### 3.2.1 Objective Function

The objective function consists of two parts. The first part is the total electricity cost which can be expressed as

$$J_{\text{elec}} = \sum_{t=1}^{NH} P_t^{\text{grid}} e_t \Delta T. \tag{3.2}$$

Substitute the result of (3.1) into (3.2), we have

$$J_{\text{elec}} = \sum_{t=1}^{NH} \left( \sum_{j=1}^{NB} P_{j,t}^{\text{hvac}} + \sum_{j=1}^{NB} \sum_{e=1}^{NE_j} \left[ P_{j,e,t}^{\text{ev,c}} - P_{j,e,t}^{\text{ev,d}} \right] \right) e_t \Delta T. \tag{3.3}$$

The second part of the objective function is the discomfort cost. It is assumed that each household $j$ will inform its preferred temperature $T_{j,t}^{\mathrm{d}}$ for each time slot $t$ when the house is occupied and a maximum acceptable temperature deviation $\delta_{j,t}$ to the aggregator [1]. Then, the aggregator controls the power supplied to HVAC system so that the temperature lies within the acceptable range $[T_{j,t}^{\mathrm{d}} - \delta_{j,t}, T_{j,t}^{\mathrm{d}} + \delta_{j,t}]$ in each occupied house $j$. The closer to the desired temperature, the more comfortable users would be. There is no indoor temperature constraint for the period when the house is unoccupied. The power supplied to an HVAC system at time slot $t$ will decide the temperature at time slot $t+1$, which consequently affects user comfort at time slot $t+1$ but not time slot $t$. Therefore, we define the discomfort cost function as

$$J_{\mathrm{discomfort}} = \sum_{t=1}^{NH} \sum_{j=1}^{NB} w_{j,t} a_{j,t+1} |T_{j,t+1}^{\mathrm{in}} - T_{j,t+1}^{\mathrm{d}}|, \tag{3.4}$$

where $a_{j,t+1}$ represents the occupancy status of house $j$ at time slot $t+1$. If $a_{j,t+1} = 0$, the discomfort cost for house $j$ at time slot $t+1$ is equal to zero regardless of the indoor temperature at that time slot because the house is not occupied. The weighting factor $w_{j,t}$ can be viewed as the price ($\$$) that aggregator has to pay household $j$ at time $t$ when the temperature in house $j$ deviates 1°C from the desired temperature in each time slot. The value of $w_{j,t}$ will influence the optimal solution of the underlying optimization problem. The objective function, which is the sum of electricity cost and the discomfort cost, can be written as follows:

$$J_{\mathrm{tot}} = \sum_{t=1}^{NH} \left( \sum_{j=1}^{NB} P_{j,t}^{\mathrm{hvac}} + \sum_{j=1}^{NB} \sum_{e=1}^{NE_j} \left[ P_{j,e,t}^{\mathrm{ev,c}} - P_{j,e,t}^{\mathrm{ev,d}} \right] \right) e_t \Delta T + \sum_{t=1}^{NH} \sum_{j=1}^{NB} w_{j,t} a_{j,t+1} |T_{j,t+1}^{\mathrm{in}} - T_{j,t+1}^{\mathrm{d}}|$$

$$\tag{3.5}$$

We are now ready to describe all constraints for the considered optimization problem.

## 3.2.2 Thermal Constraints

### 3.2.2.1 Thermal Dynamics Model

The thermal dynamics model for buildings has been presented in Section 2.1. It can be seen from equation (3.6), for each building, the indoor temperature at the next time slot is

---

[1]In the proposed control scheme, we consider only customers who agreed to participate in this scheme. For those customers who do not participate in the control scheme, their load will be treated as normal load and uncontrollable.





determined by the current indoor temperature $(T_{j,t}^{\text{in}})$, the current outdoor temperate $(T_t^{\text{a}})$, the solar radiation power $(Q_{j,t}^{\text{sr}})$, and the output power of the HVAC system $(P_{j,t}^{\text{hvac,out}})$ at the current time slot. The output power is related to the power supplied to the HVAC system as $P_{j,t}^{\text{hvac,out}} = \eta_j P_{j,t}^{\text{hvac}}$ where $\eta_j$ is the coefficient of performance (COP) of the HVAC system in building $j$. The discrete time thermal dynamic model (3.6) represents one constraint of the considering optimization problem.

$$T_{j,t+1} = A_j^{\text{d}} T_{j,t} + B_j^{\text{d}} U_{j,t}, \quad \forall j, \ t$$
$$T_{j,t}^{\text{in}} = C_j^{\text{d}} T_{j,t}, \qquad \forall j, \ t \tag{3.6}$$

where $T_{j,t} = [T_{j,t}^{\text{in}}, \ T_{j,t}^{\text{m}}, \ T_{j,t}^{\text{e}}]'$ and $U_{j,t} = [T_t^{\text{a}}, \ \Phi_t, \ \sigma_j \eta_j P_{j,t}^{\text{hvac}}]'$. The calculation of matrices $A_j^{\text{d}}$, $B_j^{\text{d}}$, and $C_j^{\text{d}}$ was presented in Section 2.1.

### 3.2.2.2 Temperature Constraints

Each household informs its desired temperature to the aggregator. Then, the aggregator controls power supplied to the HVAC system in the house at each time slot to keep the indoor temperature as close as possible to the desired temperature. The indoor temperature requirement for each house is expressed as

$$a_{j,t}|T_{j,t}^{\text{in}} - T_{j,t}^{\text{d}}| \le \delta_{j,t}, \tag{3.7}$$

for $j = 1, 2, \ldots, NB$ and $t = 2, \ldots, NH + 1$. There is no temperature requirement when a house is not occupied. Note that the power provided to an HVAC system in the current time slot will affect the indoor temperature in the next time slot, so the temperature constraint is only applied from the second time slot.

### 3.2.2.3 HVAC Power Constraints

The power supplied to an HVAC system cannot be negative and it cannot take values greater than the heater/AC's power rating. Therefore, we have

$$0 \le P_{j,t}^{\text{hvac}} \le P_j^{\text{hvac,max}}, \tag{3.8}$$

for $j = 1, 2, \ldots, NB$ and $t = 1, \ldots, NH$.

In our proposed system model, each household needs to report its desired temperature at each time slot during which the household is occupied as well as the level of discomfort (i.e., parameters $\delta_{j,t}$) to the aggregator to determine the optimal control solution. In practice, if





a particular user does not wish to report its desired temperature in the occupied time slots to the aggregator then the aggregator can simply choose a typical temperature value for this household to calculate the optimal solution.

### 3.2.3 SOC and Charging Power Constraints

For EVs, we need to model the characteristics and the travel patterns for each EV. In particular, we are interested in the battery capacity ($kWh$), the charging rated power ($kW$), the travel efficiency ($kWh/km$), and the charging type. These properties can be retrieved from the manufacturer's website. The travel pattern of each EV can be described by the number of trips per day, the starting and ending times, and the travel distance of each trip. A trip is defined as the time period between the instants when the EV leaves and arrives home. This information is related to user traveling schedule, which can be sent by users to the aggregator before the operating day. In Section 3.3, we use the real-world travel pattern data from the 2009 National Household Travel Survey [89] to build travel patterns used to obtain numerical results.

At the current stage of EV technologies, the vehicle-to-grid (V2G) ability is not readily available. Moreover, car and battery manufacturers consider that V2G operation would essentially void a car battery's warranty. However, the V2G ability is expected to be widely available for next generations of EVs [90]. Therefore, the proposed optimization framework is valuable in the future (not in the near future though) when the V2G technology is more matured. Last but not least, we assume that the degradation cost of battery charging/discharging activities is negligible in this Chapter. A more detailed battery degradation cost due to V2G operation is subject of future work. A simple model of battery degradation cost can be found in Chapter 4.

#### 3.2.3.1 SOC Dynamics

Assume that each EV $e$ of house $j$ can take several trips during the optimization period (e.g., one day). Let $t_{j,e,l}^{(1)}$ and $t_{j,e,l}^{(2)}$ be the time slots when EV $e$ of house $j$ leaves and arrives home for trip $l$, respectively. Then, we have following constraints





$$SOC_{j,e,t+1} = SOC_{j,e,t} + \frac{\eta_{j,e}^{c} P_{j,e,t}^{\mathsf{ev,c}} \Delta T}{E_{j,e}^{\mathsf{cap}}} - \frac{P_{j,e,t}^{\mathsf{ev,d}} \Delta T}{\eta_{j,e}^{\mathsf{d}} E_{j,e}^{\mathsf{cap}}}, \text{ if } t \notin [t_{j,e,l}^{(1)}, t_{j,e,l}^{(2)}), \quad \forall j, t, e, l \qquad (3.9)$$

$$SOC_{j,e,t+\Lambda_{j,e,l}} = SOC_{j,e,i} - \frac{d_{j,e,l} * m_j}{E_{j,e}^{\mathsf{cap}}}, \text{ if } t = t_{j,e,l}^{(1)}, \quad \forall j, t, e, l \qquad (3.10)$$

$$SOC_{j,e,t_{j,e,l}^{(2)}} \leq SOC_{j,e,t} \leq SOC_{j,e,t_{j,e,l}^{(1)}}, \text{ if } t \in [t_{j,e,l}^{(1)}, t_{j,e,l}^{(2)}], \quad \forall j, t, e, l. \qquad (3.11)$$

Here, the SOC for EV $e$ of house $j$ changes according to the charging and discharging powers when it parks at home (3.9) and the difference of SOCs at leaving and returning home instants accounts for the energy usage in driving (3.10). Equation (3.11) ensures that the SOC level is non-increasing when an EV travels.

### 3.2.3.2 SOC Constraints

To maintain long lifetime of battery, an EV should maintain its battery level within a certain range that is recommended by its manufacturer [90]. Therefore, we impose the following constraints

$$SOC_{j,e}^{\mathsf{min}} \leq SOC_{j,e,t} \leq SOC_{j,e}^{\mathsf{max}}, \quad \forall j, \ e, \ t \qquad (3.12)$$

where $SOC_{j,e}^{\mathsf{min}}$ and $SOC_{j,e}^{\mathsf{max}}$ denote the minimum and maximum recommended SOCs for EV $e$ of house $j$.

### 3.2.3.3 Charging and Discharging Constraints

We assume that an EV is only charged or discharged when it is parked at home. Moreover, EVs are connected to home chargers as soon as they arrive home. Therefore, constraints on charging and discharging power applied to only time slots when an EV is parked at home as

$$0 \leq P_{j,e,t}^{\mathsf{ev,c}} \leq b_{j,e,t} P_{j,e}^{\mathsf{ev,c,max}}, \quad \forall j, \ e, \ t$$
$$0 \leq P_{j,e,t}^{\mathsf{ev,d}} \leq b_{j,e,t} P_{j,e}^{\mathsf{ev,d,max}}, \quad \forall j, \ e, \ t. \qquad (3.13)$$

where $b_{j,e,t}$ represents the availability of EV $e$ of house $j$ at home during time slot $t$, $P_{j,e}^{\mathsf{ev,c,max}}$ and $P_{j,e}^{\mathsf{ev,d,max}}$ denote the maximum charging and discharging limits, respectively. From these constraints, the charging and discharging powers for each EV $e$ of household $j$ (i.e., $P_{j,e,t}^{\mathsf{ev,c}}$ and $P_{j,e,t}^{\mathsf{ev,d}}$) are equal zero if the EV is not at home (i.e., as $b_{j,e,t} = 0$).





### 3.2.4 Grid Constraints

If it is not allowed to sell EV discharging energy back to the main grid, the energy imported from the grid in each time slot must be non-negative and it must be upper-bounded by some predetermined limit. Hence, we have

$$0 \leq P_t^{\text{grid}} \leq P_t^{\max}, \quad \forall t \tag{3.14}$$

or

$$0 \leq \sum_{j=1}^{NB} P_{j,t}^{\text{hvac}} + \sum_{j=1}^{NB} \sum_{e=1}^{NE_j} [P_{j,e,t}^{\text{ev,c}} - P_{j,e,t}^{\text{ev,d}}] \leq P_t^{\max}, \quad \forall t \tag{3.15}$$

where $P_t^{\max}$ is the maximum power that can be imported from the grid, which can be a contracted amount between the aggregator and the grid, or a particular parameter capturing grid conditions over time. In contrast, if the selling EV discharging energy service is allowed and the maximum power that can be sold back to the main grid is equal to $P_t^{\max}$. Then, we have the following constraint

$$-P_t^{\max} \leq \sum_{j=1}^{NB} P_{j,t}^{\text{hvac}} + \sum_{j=1}^{NB} \sum_{e=1}^{NE_j} [P_{j,e,t}^{\text{ev,c}} - P_{j,e,t}^{\text{ev,d}}] \leq P_t^{\max}, \quad \forall t \tag{3.16}$$

For simplicity, we assume that the selling back electricity price is equal to the buying electricity price. In summary, we can formulate the EV charging and HVAC scheduling to minimize the cost function $J_{\text{tot}}$ given in (3.5) as

$$\min \sum_{t=1}^{NH} \left( \sum_{j=1}^{NB} P_{j,t}^{\text{hvac}} + \sum_{j=1}^{NB} \sum_{e=1}^{NE_j} \left[ P_{j,e,t}^{\text{ev,c}} - P_{j,e,t}^{\text{ev,d}} \right] \right) e_t \Delta T + \sum_{t=1}^{N} \sum_{j=1}^{NB} w_j a_{j,t+1} |T_{j,t+1}^{\text{in}} - T_{j,t+1}^{\text{d}}|$$

subject to

$$\text{constraints } (3.7) - (3.13)$$

$$\text{constraints } (3.15), \text{ if no V2G}$$

$$\text{constraints } (3.16), \text{ if V2G} \tag{3.17}$$

where the optimization variables are $P_{j,e,t}^{\text{ev,c}}$, $P_{j,t}^{\text{hvac}}$, and $P_{j,e,t}^{\text{ev,d}}$. Despite of the absolute term in the objective function of our model, this optimization problem can be reformulated to an equivalent linear program with some auxiliary variables [91] as described in Section 2.2.2. Thus, the aggregator can easily calculate and implement its optimal solution upon collecting all required information.





It can be observed that we do not impose constraints in the optimization problem (3.17) to prevent any EV $e$ of house $j$ from charging and discharging simultaneously at any time slot $t$ (i.e., $P_{j,e,t}^{ev,c}$ and $P_{j,e,t}^{ev,d}$ for any EV $e$ of house $j$ are both positive at the same time slot $t$). In fact, this is not needed since the optimal solution of (3.17) always satisfies these hidden constraints.

### 3.2.5 Extensions to Consider Modeling Uncertainties and Other Loads

In the above formulation, we have assumed that all modeling parameters such as outdoor temperature, household occupancy pattern, and EV travel pattern are known without errors and the thermal dynamics model is perfect. In practice, they have to be estimated with potential errors. We can employ the Model Predictive Control (MPC) technique to tackle these estimation uncertainties [92], which can be implemented as follows. The MPC controller solves the minimization problem (3.17) for the prediction horizon $N_0$ from current time slot $t$ to time slot $t + N_0$ with assumption that estimated parameters are certain ones (i.e., no estimation errors). The uncertainties are compensated by refinement and update of the prediction at each time step. The sequences of control variables such as power consumption of HVAC and EVs are calculated for the whole prediction horizon; however, the controller applies only the control action for the first time slot. The MPC controller repeats the process at next time step, solving a new optimization with the most updated data for the new time horizon shifted one step forward.

We have only considered EVs and HVAC systems in our proposed optimization framework so far. However, integration of other types of residential loads into this framework is possible. Moreover, extension of our system model to consider distributed renewable energy sources is also possible. Here, the MPC technique can be employed again to tackle the uncertainty due to the intermittent nature of the underlying renewable sources (e.g., wind or solar energy).

## 3.3 Numerical Results

We present numerical results to illustrate the desirable performance of the proposed framework. We assume that the outdoor temperature and solar irradiance can be predicted perfectly. The temperature data is taken from *Weather Underground* website [93], and the solar irradiance data is taken from the Renewable Resource Data Center (RReDC) website [94]. Solar radiation power contributes to increase the indoor house temperature; therefore,





it results in more cooling energy needed in the summer and less heating energy needed in the winter. Moreover, the solar irradiance is low during the winter months and high during summer months as we can observe in Fig. 3.2(d). Hence, the decreasing amount of heating energy required in the winter is relatively small compared to the increasing amount of cooling energy needed in the summer.

For the electricity price data, we use day-ahead pricing data retrieved from PJM [95]. Simulation data of weather conditions and electricity price are taken at the same location. We will first evaluate the performance of our control scheme for a single-house scenario. Then, we investigate the benefits of applying the control strategy in the multiple-house setting. Results for the single-house scenario are presented to reveal insights into the interaction among the HVAC system, EV, pricing, and temperature patterns.

When solving the joint scheduling optimization for EVs and HVAC systems we set $\sigma_j = -1$ and $\sigma_j = 1$ for all $j$ corresponding to the summer and winter time, respectively. The optimization period is one day with 24 time slots each of which is one hour ($NH = 24$, $\Delta T = 1$). Fig. 3.2(a) shows the day-ahead electricity prices of typical summer and winter weekdays, which are used to obtain numerical results. Three different temperature profiles for summer days (*very hot, hot, mild*) and winter days (*very cold, cold, mild*) are considered to represent the diversity of weather conditions, as shown in Fig. 3.2(b) and Fig. 3.2(c), respectively. The average hourly solar irradiance profiles for the summer and winter cases used in the simulation are shown in Fig. 3.2(d). We use CVX software [85] to solve the proposed optimization problem.

### 3.3.1 Single-house Scenario

We analyze the performance due to our proposed optimal scheme to the single-house scenario. For simplicity, we assume that the considering residential house has only one EV and one HVAC where the varying ownership aspect will be captured later in Section V.B for the multiple-house case. The housing thermal parameters including $R_j^{\mathsf{a}}$, $R_j^{\mathsf{m}}$, $R_j^{\mathsf{e}}$, $R_j^{\mathsf{ea}}$, $C_j$, $C_j^{\mathsf{m}}$, and $C_j^{\mathsf{e}}$ are taken from [41]. We assume that the house is equipped with a heat pump which can be operated in both heating and cooling modes. The parameters of the heat pump are set as follows: power rating $P_j^{\mathsf{hvac,max}} = 4$ kW and HVAC coefficient of performance (COP) $\eta_j = 3$. We consider Nissan Leaf EVs whose specifications are obtained from [90] with the following parameter setting: battery capacity of 24 kWh; maximum charging and discharging powers are set equally to 6 kW [90]; charging and discharging efficiency factors are both set equal to 0.9; travel efficiency is 0.316 kWh/mile; and the maximum and minimum SOC are





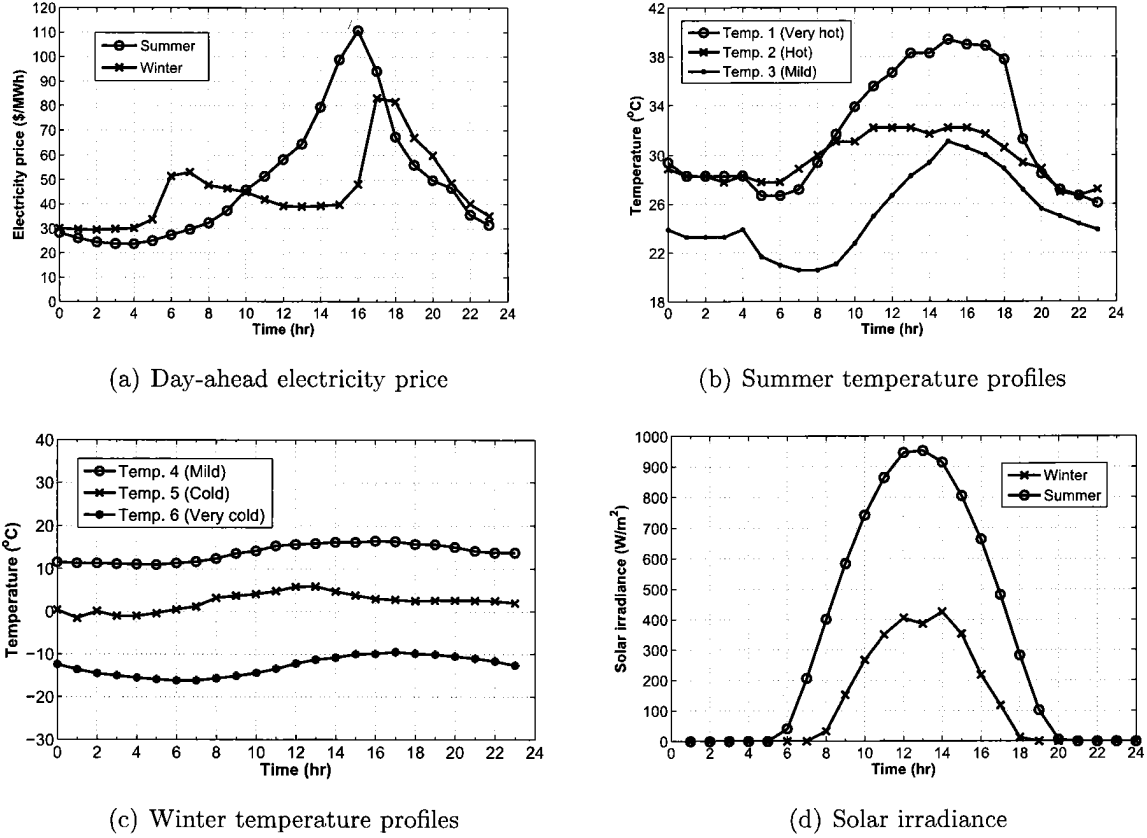

(a) Day-ahead electricity price      (b) Summer temperature profiles

(c) Winter temperature profiles      (d) Solar irradiance

**Figure 3.2:** Electricity price and weather profiles

0.9 and 0.2, respectively. The initial SOC of the EV is set equal to 0.5. To obtain numerical results for the single-house scenario, we simply set $NB = 1$ in all related constraints and quantities.

We assume that the EV's owner leaves home at 8 A.M and comes back at 5 P.M. Driving distance is assumed to be 32 miles, which is the average daily travel distance in US [28, 89, 96]. This is a typical driving pattern [27] in US, which is used to obtain the numerical results in several scenarios below. However, other different driving patterns are also examined where we will investigate the impacts of varying departure time, arrival time, and travel distance on the optimal solution. In addition to the above parameter settings for HVAC system and EV, the power limit $P_t^{\text{max}}$ is set equal to 25 kW. The desired indoor temperature in summer days and winter days are 23°C and 21°C, respectively. Initial temperature at 0 A.M is assumed to be equal to the desired indoor temperature. Assuming that the considered house is occupied all day, so the desired indoor temperature is equal to 23°C (summer) or 21°C (winter) at





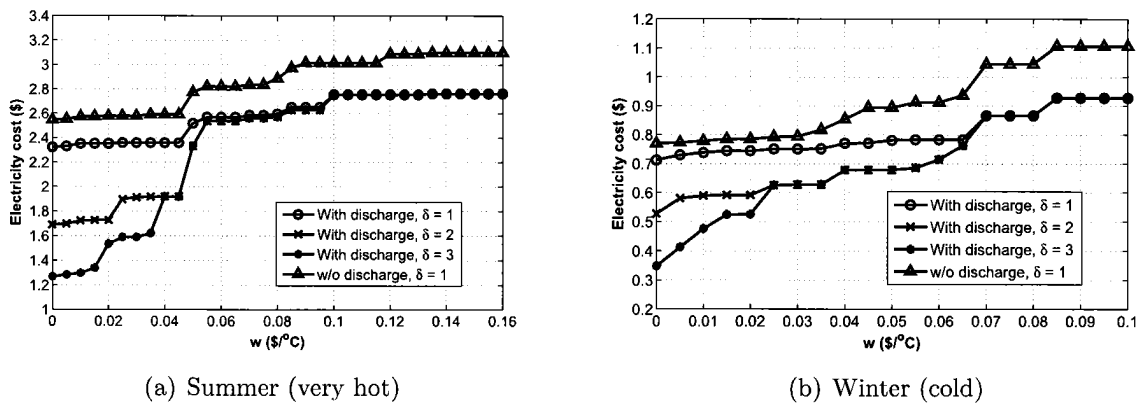

(a) Summer (very hot)          (b) Winter (cold)

**Figure 3.3:** Impacts of parameters $w$ and $\delta$ on electricity cost (No V2G )

every time slot.

Fig. 3.3(a) and Fig. 3.3(b) show the impacts of parameters $\delta$ and $w$ on the total electricity cost in a summer day and a winter day, respectively. The weighting factor $w$ is varied in a certain interval, and $\delta$ takes one of three values: 1°C, 2°C, and 3°C. To enforce the stricter user comfort requirement, we would choose a higher value for $w$ and a smaller value for $\delta$. These figures show that the electricity cost increases as $w$ increases. This is intuitive since the cost of electricity increases with stricter user comfort requirement. Moreover, for small values of $w$ (i.e., $w < 0.04$ \$/°C for the summer day, and $w < 0.025$ \$/°C for the winter day), the electricity cost decreases as $\delta$ increases. This is because small values of $w$ allow the indoor temperature to deviate more significantly from the preferred value to save electricity cost, especially for large values of $\delta$. However, as $w$ becomes sufficiently large (i.e., $w > 0.1$ \$/°C for the summer day, and $w > 0.07$ \$/°C for the winter day), the electricity costs corresponding to the three different values of $\delta$ are the same. This is because sufficiently high values of penalty value $w$ results in the temperature being close to the desired value for the whole day. These figures also show that the electricity cost due to optimal control without discharging is much higher than that exploiting EV discharging capability. This result confirms the great benefits of exploiting interactions among EVs and between EVs and HVAC systems. To obtain numerical results for the optimal control without discharging, we simply set maximum discharging power to zero in our problem formulation.

Due to the space constraint, we consider only the summer case to present other numerical results in the following. Fig. 3.4(a) illustrates the indoor temperature variation over time for $\delta = 2$°C and different values of $w$. This figure confirms our observation from Fig. 3.3(a)





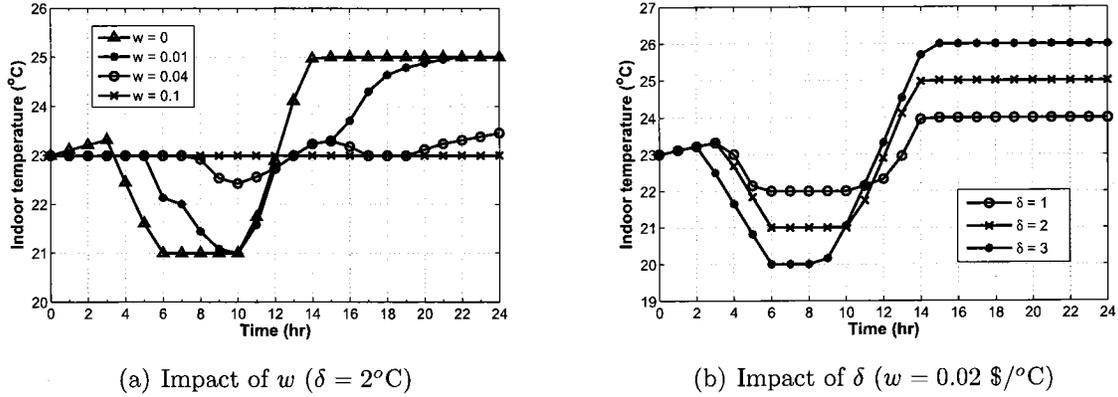

(a) Impact of $w$ ($\delta = 2°C$)  (b) Impact of $\delta$ ($w = 0.02$ \$/°C)

**Figure 3.4:** Impacts of parameters $w$ and $\delta$ on indoor temperature (No V2G)

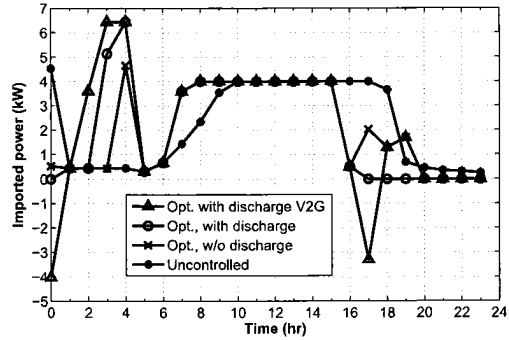

**Figure 3.5:** Imported power in different schemes

where higher values of $w$ reduce the fluctuation of indoor temperature around its preferred value. In Fig. 3.4(b) we plot the indoor temperature over time for a fixed $w$ and different values of $\delta$. This figure again indicates that the indoor temperature oscillates more around the preferred value as $\delta$ increases. Furthermore, we can observe that the indoor temperature tends to be low during low price hours and becomes higher during high price hours. This result shows that our proposed power scheduling scheme will precool buildings during low price hours (i.e., by consuming more energy) so that it allows us to reduce the energy consumption of buildings during high price hours.

Fig. 3.5 shows the power imported from the grid under our control scheme exploiting EV discharging capability compared to optimal solution without discharging and the uncontrolled scheme in the summer day. For the uncontrolled scheme, EV charging occurs at midnight when EV is plugged, regardless of electricity price. The charging terminates





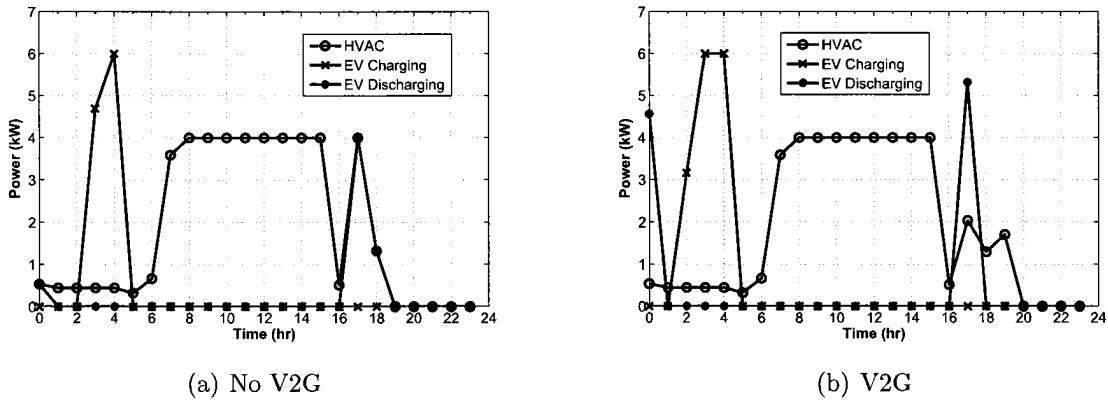

(a) No V2G          (b) V2G

**Figure 3.6:** Power profile (very hot day, $w = 0.01$ \$/°C, $\delta = 2$°C)

when the EV meets the energy consumption requirement for the day. For fair comparison, the energy charged to EV is chosen so that the remaining energy in the EV battery at the end of the day equal to the energy in the controlled case. In addition, for the uncontrolled case, the HVAC is controlled by a thermostat to keep the indoor temperature equal to the desired temperature at every time slots. It can be observed that our optimal scheme with discharging reduces significantly the load during peak hours (from 2 P.M to 9 P.M) when the electricity price is very high (cf. Fig. 3.2(a)). The negative value of imported power in the V2G case represents the power selling back to the grid. To supply energy for the HVAC system, the EV discharges its remaining battery right after it arrives home as indicated in Figs. 3.6(a), 3.6(a). Moreover, during high-price hours the amount of the EV discharging power for the V2G case in Fig. 3.6(b) is much higher than the amount of the EV discharging power without V2G service in Fig. 3.6(a). EV charging occurs at time slots when electricity prices are low (from 3 A.M to 4 A.M).

In Figs. 3.7(a), 3.7(b), we present the impacts of different temperature profiles on electricity cost saving compared to the uncontrolled scheme in a summer day. It can be observed that the cost saving decreases with increasing $w$ since larger values of $w$ reduces the flexibility in controlling HVAC consumption. Also, the absolute cost saving in dollars is larger for a hot or very hot day than that for a mild day. This is because more EV discharging energy to the HVAC system would be expected in a hot or very hot day, which translates into more significant cost saving. Fig. 3.7(b) shows that for a small value of $w$ (e.g., less than 0.04 \$/°C), the cost saving of more than 25% can be achieved in a mild or hot day where the





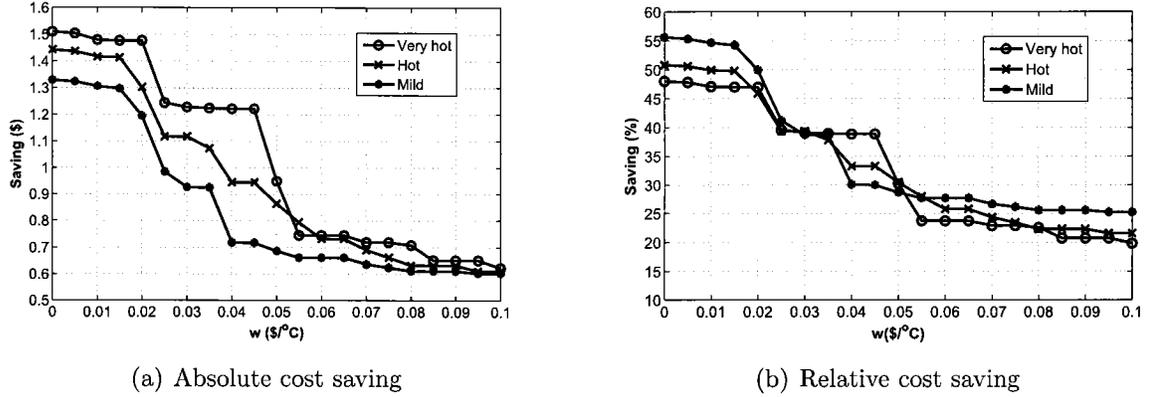

(a) Absolute cost saving            (b) Relative cost saving

**Figure 3.7:** Impact of temperature profiles and $w$ on cost saving (No V2G, $\delta = 2^oC$)

relative cost saving is calculated as

$$\text{Saving}(\%) = \frac{\text{Cost of uncontrolled case - Optimal cost}}{\text{Cost of uncontrolled case}} 100.$$

We illustrate the absolute and relative cost saving compared to the uncontrolled scheme versus users' arrival and departure times for different summer temperature profiles in Figs. 3.8(a)-3.8(d), respectively. These figures show that the absolute cost and the relative cost saving decrease with increasing users' arrival time (departure time is fixed at 8 A.M) while it increases with increasing users' departure time (arrival time is fixed at 5 P.M). These results can be interpreted as follows. The relative cost saving would increase if EVs are available at home for a longer duration per day. This is because by connecting with the power grid longer, EVs can charge their batteries during off-peak hours and discharge energy to supply the HVAC system in on-peak hours more efficiently. However, EV parking time at home is directly related to users' arrival and departure times. Also, it is easy to recognize that by using V2G service, more cost saving would be achieved.

To investigate the impact of travel distance on the cost saving of the proposed scheme compared to the uncontrolled one, we fix the departure time (8 A.M.) and the arrival time (5 P.M.) and vary the travel distance of the EV. The EV battery SOC when it returns home depends on its energy consumption which, in turn, depends on the travel distance. Moreover, the electricity price is high around 5 P.M.; therefore, the higher SOC when the EV gets home, the more energy can be discharged from the EV to supply power to the HVAC system, which consequently results in larger cost saving. The numerical results in Figs. 3.9(a), 3.9(b) confirm this point by showing that the cost saving decrease as the travel distance increases.





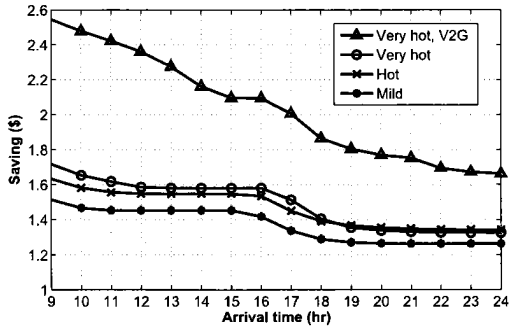

(a) Varying arrival time (absolute)

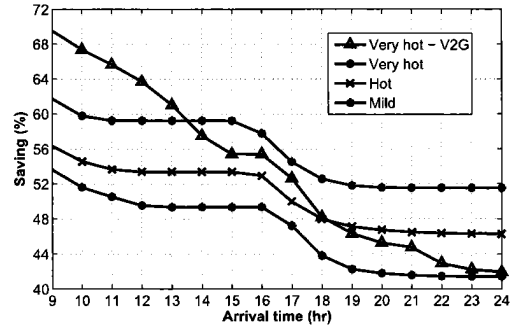

(b) Varying arrival time (relative)

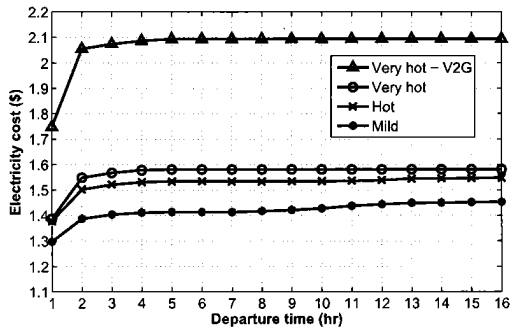

(c) Varying departure time (absolute)

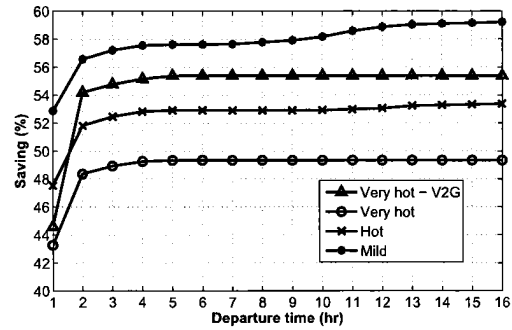

(d) Varying departure time (relative)

**Figure 3.8:** Impact of departure time and arrival time on cost saving ($\delta = 2^o$C, $w = 0.01$ $/$^o$C)

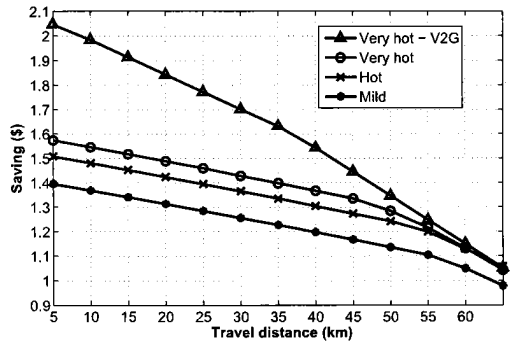

(a) Absolute saving

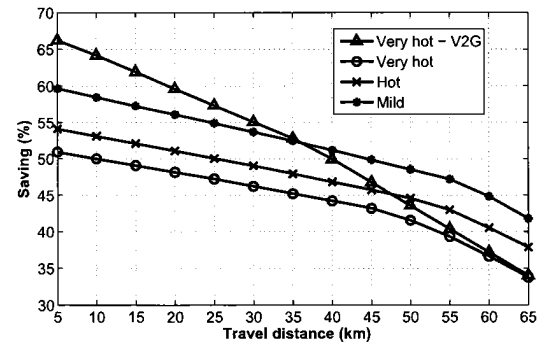

(b) Relative saving

**Figure 3.9:** Impact of travel distance on cost saving (very hot, $\delta = 2^o$C, $w = 0.01$ $/$^o$C)





### 3.3.2 Multiple-house Scenario

In the previous section, we have shown the strength of our proposed control scheme for the single-house scenario compared to the optimal scheme without discharging and the uncontrolled one. In this section, we will demonstrate that, it is even more cost-efficient if we apply our proposed control scheme to manage a group of households. Toward this end, we consider a community of 100 households ($M = 100$). We also assume that all EVs are Nissan Leaf whose specifications such as battery capacity, maximum charging/discharging power are described in the previous section. We take building thermal parameters from [41] as mean values for thermal parameters of houses in the community. Each thermal parameter (resistance and capacitance) of each house is chosen to be uniformly distributed in the interval of +/- 20% around the mean value to represent the diversify of houses in the community. In practice, the size of AC/heater units would be chosen based on the size and shape of the building. To capture the variety of HVAC systems, we assume that the power rating and COP of HVAC units in the community are uniformly distributed in [4, 6] kW and [2.5, 3.5] intervals. We assume that all households are occupied all day and the desired indoor temperature for all households is 23ºC in the summer. The initial indoor temperature is set randomly in [22ºC, 24ºC] interval and the maximum imported power from the grid is set equal to 1 MW. The initial SOC of EVs are chosen to be uniformly distributed in the range [$SOC^{min}$, $SOC^{max}$].

The travel patterns of EVs in the community are randomly generated based on statistical data from National Household Travel Survey (NHTS) data set, which collects daily travel information of households in the U.S. [89]. For simplicity, only the departure time of the first trip and the arrival time of the last trip are taken into account even though our proposed model can cover multiple trips per day. This assumption can be justified because if an EV comes back home for a short time during the day, its available time for charging/discharging between the trips is small and the benefit due to the energy exchange would be insignificant. We choose the departure times for different EVs randomly according to a normal distribution with the mean of 7 A.M. and the standard deviation of 2 hours. The arrival time is drawn randomly according to another normal distribution with the mean of 6 P.M. and the standard deviation of 2 hours. These parameter settings were suggested by [97], which were established by using the data set given in [89]. Daily travel distance also follows a log-normal distribution with the mean of 32 miles and a standard deviation of 24 miles [96]. Based on EV ownership information [89], we assume that there are 9 households which have no EV, 32 households each of which has 1 EV, 36 households each of which has 2 EVs, 12 households with 3





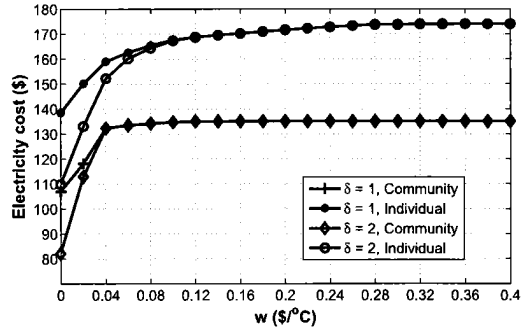

**Figure 3.10:** Comparison optimal electricity cost under community-based optimal and individual-based optimal solutions (No V2G)

EVs each, and 11 households each of which has 4 EVs. Therefore, there are 184 EVs under consideration. There are about 35% of vehicles do not travel all day according to [25, 28, 89]; therefore, we set 64 EVs out of the 184 EVs to be available at home all day. For remaining EVs, their travel patterns (departure time, arrival time, and travel distance) follow the normal and log-normal distributions as described above.

To ease the analysis and presentation, we assume that all households choose the same value of $w_j = w$ and same the value of $\delta_{j,t} = \delta$ in our model. We evaluate the performance for two cases where the proposed optimization framework is applied for the whole community with 100 households and for individual households, respectively. The results corresponding to these two cases are indicated as "Community" and "Individual" in Figs. 3.10, 3.11(a), respectively. Fig. 3.10 shows the optimal costs for both control schemes in the *very hot summer day*. It can be observed that the total electricity cost is reduced quite significantly when we optimize the energy usage for the whole community compared to the case when each household optimizes its energy consumption separately. The performance gain is about 20%.

Fig. 3.11(a) plots the power imported from the grid for *a hot summer day* and $w = 0.02$ $/°C. Fig. 3.11(b) illustrates the total power consumption and charging/discharging due to HVACs and EVs in the community. As can be seen, the total energy imported from the grid in the community-optimization scheme is zero during peak hours (1 P.M to 6 P.M). When each household applies our optimal scheme separately, the total energy imported from the grid is non-zero for several high-price hours. This "demand response" effect achieved by the proposed community-optimization scheme is very desirable since it helps reduce the peak demand in on-peak hours. In addition, Fig. 3.11(b) shows that by exploiting EV discharging





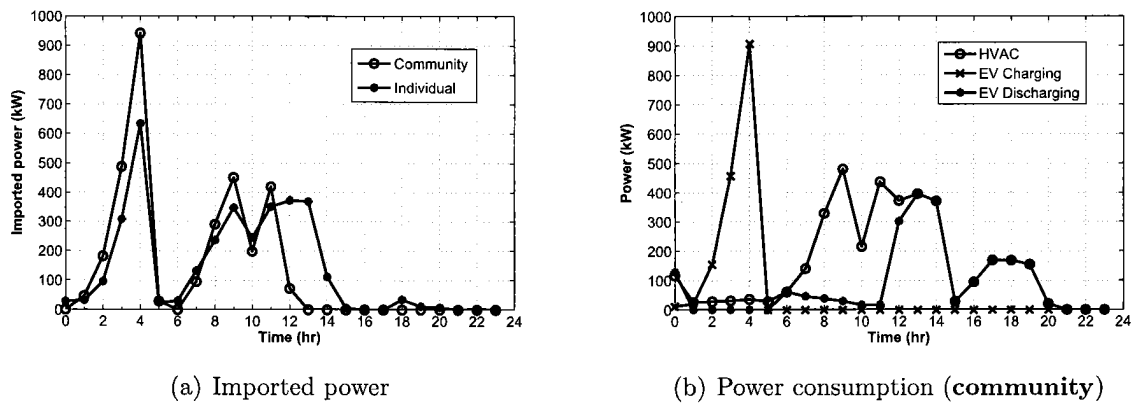

(a) Imported power        (b) Power consumption (**community**)

**Figure 3.11:** Imported power and power consumption, charging/discharging (very hot day, no V2G, $w = 0.02$ \$/°C, $\delta = 2$°C)

capability we can reduce the demand and save the electricity cost during high-price hours.

## 3.4 Summary

This chapter presented a unified framework to jointly optimize the EV and home energy scheduling considering user comfort preference. The proposed design captures different key modeling aspects including thermal dynamics, EV travel, user occupancy patterns, as well as operational constraints of the HVAC system and EVs. Extensive numerical results were shown to demonstrate the impacts of different parameters on the electricity cost, the significant gain achieved by the proposed solution, and the benefits due to optimization of EV and home energy scheduling for multiple houses in a residential community.



# Chapter 4

# Optimal Energy Bidding and Scheduling for Microgrids

The joint scheduling optimization of EVs and HVACs considered in the previous chapter was formulated as a deterministic optimization problem. In reality, many system parameters are uncertainties. In this chapter, we consider a different application scenario where the flexible HVAC load is utilized to compensate for the variability of renewable energy generation. Specifically, we study an energy bidding and scheduling problem for a microgrid (MG) which participates in the electricity market where the MG is assumed to be a price-taker. The underlying optimization problem is formulated using a two-stage stochastic program. We will present briefly the concept of MG, review the operation of HVAC, and the electricity market model followed by the solution approach, which employs the stochastic programming technique. The objective function and relevant constraints of the proposed optimization problem are then described. Finally, the advantages of the proposed economic model is illustrated through extensive numerical results with a series of sensitivity analyses.

## 4.1 System Model and Modeling Approach

### 4.1.1 System Model

#### 4.1.1.1 MG Components

We consider a large-scale MG that consists of several renewable generating units, conventional units, a number of buildings with associated loads, and an optional battery storage facility. The renewable generating units include solar panels and wind turbines. In this





study, conventional generating units (DG) refer to non-renewable generating units such as microturbines, fuel cells, and diesel generators.[1] The MG loads are divided into two separate types, namely HVAC and non-HVAC loads. In reality, this kind of MG is very popular (e.g., *energy cooperative* model [98]).

In general, the MG aggregator desires to maximize the utilization of renewable energy generation. The shortened and excess amount of energy required to balance the local load can be accommodated by trading with the main grid through the Point of Common Coupling (PCC) [58] or by running conventional units. The MG aggregator will make decisions on purchasing electricity from the market or running local conventional units depending on various factors such as electricity price, the states of conventional units, and the marginal cost of operating conventional units. We consider the energy scheduling and bidding problem for MG in discrete time slots, denoted by $t$ in our model, over a scheduling period of NH time slots.

### 4.1.1.2 HVAC Operation

HVAC systems are typically controlled by the thermostats to maintain indoor temperature at preferred setpoints. Users can choose desirable temperature setpoints for different occupancy statuses (e.g., being at home, away, and slipping). Users would feel the most comfortable if the indoor temperature is at the preferred setpoint. However, they can tolerate a certain small deviation of the indoor temperature from the setpoint. The larger the deviation is, the less comfortable users would feel. In this work, we assume that the indoor temperature must be maintained to be in the comfort range (e.g., [21°C, 25°C]).

Moreover, indoor temperature in a particular time slot depends on the amount of scheduled power and the temperature in the previous time slot due to building thermal inertia. Therefore, it would be beneficial if the HVAC system consumes more energy when the electricity price is low or when the amount of generated renewable energy is high to precool or preheat buildings (for the summer and winter, respectively) and reduce the power consumption in opposite cases. Thus, the HVAC power consumption should be scheduled economically while maintaining the indoor temperature within the required comfort range. By allowing the MG aggregator to control the operation of HVAC systems in their buildings, significant cost saving may be achieved based on which users of the MG could receive some saving in their electric bills.

---

[1]In reality, conventional units are often refereed to as thermal generators.





### 4.1.1.3 Market Model

The MG is assumed to be a price-taker in the electricity market. During the time slots where the local power generation is surplus, the MG would sell its power to the main grid. In opposite, if the local generation is not enough to meet its local load, the MG would need to buy electricity from the main grid. Everyday the MG has to submit its hourly bids to the day-ahead market several hours before physical power delivery [31]. The MG bids include both selling and buying electricity bids. Also, the MG can participate in the real-time electricity market to supplement for any power deviation from the day-ahead schedule. It is the common practice that the MG offers selling bids at a very low price (e.g, normally set to 0 $/MWh [31, 33, 95]) and buying bids at high prices to ensure that its submitted bids are always accepted in the market. The market operator is responsible for calculating the market clearing prices after collecting all offer bids and demand bids from all competitive entities in the market [44, 95]. The payment made between the MG and the market operator is calculated based on the market clearing prices.

Finding an optimal hourly bidding strategy for the MG is a challenging task due to various uncertainties in the system, which may cause a significant deviation between the scheduled power delivery and the real-time power delivery. If the MG cannot follow the day-ahead scheduled power, a penalty will be applied to the bid deviation [30, 31, 32, 33, 49]. When the available renewable output power is higher than scheduled, it is sometimes preferable to curtail the surplus renewable power to avoid a high penalty cost on the bid deviation. The thermal storage capability of buildings can help mitigate the affects of renewable energy uncertainties by increasing HVAC power consumption when renewable energy generation is higher than scheduled, and vice versa. By exploiting this aspect in HVAC power scheduling, it is, therefore, expected that the real-time power delivery will be closer to the day-ahead schedule, and the renewable energy curtailment is reduced.

## 4.1.2 Solution Approach

There are various sources of uncertainties in our proposed model, which arise from the renewable energy generation, the total non-HVAC load, the ambient temperature, and the DA and RT electricity prices. The Monte Carlo simulation is used to generate scenarios that represent these uncertain parameters based on the corresponding distribution functions [30, 31, 32, 33, 35, 36, 37, 83, 99, 100]. Scenario reduction technique is used to reduce the computational burden.





Suppose that the forecasts for uncertain parameters in the considered system model are available. Available forecasting techniques (e.g., time series, artificial neural networks, support vector machines) for wind speed, electricity prices, temperature, solar radiation and load [44, 81, 101, 102, 103, 104] can be used to attain this. In practice, the MG aggregator can obtain forecast data from a local forecasting center. For simplicity, wind speed, non-HVAC load, solar irradiance, ambient temperature, day-ahead and real-time electricity prices are assumed to follow normal distributions where the means are set equal to the forecast values and the standard deviations are 10%, 3%, 10%, 5%, 5%, and 15% of the mean values, respectively. Furthermore, we assume that the system uncertainties are independent [105]. Modeling the correlation among the uncertain parameters is beyond the scope of this thesis.

Based on the distributions of uncertainty parameters, the Monte Carlo method and Latin Hypercube Sampling technique are employed to generate 3000 scenarios with even probability (1/3000) where each scenario contains the information of the hourly load, the hourly wind speed, hourly ambient temperature, hourly solar irradiance, and the DA and RT electricity prices over the operating day. The *fast-forward* reduction algorithm is utilized to reduce the original 3000 scenarios to 15 scenarios [82, 87]. In particular, we used GAMS/SCENRED software to run the scenario reduction process [87]. In general, a larger number of scenarios results in higher computation time while a too small number of scenarios may reduce the accuracy of the results. Considering the tradeoff between computational complexity and modeling accuracy, we decided to choose 15 as the reasonable number of reduced scenarios. Some sensitivity analysis has been conducted, which confirms that the variation of the objective function is sufficiently small if larger number of reduced scenarios is chosen. sensitivity studies are not presented in this thesis.

The power scheduling and bidding problem is formulated as a two-stage stochastic program. The inputs to the underlying problem include Monte Carlo scenarios representing system uncertainties. The outputs of the optimization problem consist of the sets of first-stage decisions and the second-stage decisions. In this thesis, the first-stage decisions include the commitment statuses of all conventional units and the hourly bid quantities submitted to the day-ahead market. The second-stage decisions include the power dispatch of all generating units, the amount of involuntary load curtailment, the real-time power delivery between the MG and the main grid, and the battery charging/discharging decisions.





## 4.2 Problem Formulation

The objective function and constraints in the underlying two-stage stochastic optimization problem are described in the following. All the second-stage decision variables are denoted with a superscript $s$ representing scenario $s$.

### 4.2.1 Objective Function

Our design goal is to maximize the following objective function.

$$
\begin{aligned}
\max \quad & -\sum_{i=1}^{NG}\sum_{t=1}^{NH}(SU_{i,t}+SD_{i,t}) - \sum_{s=1}^{NS}\rho_s\sum_{i=1}^{NG}\sum_{t=1}^{NH}C_i(P_{i,t}^s) - \sum_{s=1}^{NS}\rho_s\sum_{t=1}^{NH}\pi_{j,t}\sum_{j=1}^{NB}|T_{j,t+1}^s-T_{j,t+1}^d| \\
& + \sum_{s=1}^{NS}\rho_s\sum_{t=1}^{NH}\Delta T\left\{P_t e_t^{s,\mathsf{DA}} + (P_t^s-P_t)e_t^{s,\mathsf{RT}} - \psi_t|P_t^s-P_t| - C_k^{\deg}\sum_{k=1}^{NK}(\frac{P_{k,t}^{s,\mathsf{d}}}{\eta_k^d} + \eta_k^{\mathsf{c}}P_{k,t}^{s,\mathsf{c}}) \right. \\
& \left. \qquad\qquad\qquad -V_t^{LL}LS_t^s - V_t^{\mathsf{W}}\cdot\sum_{w=1}^{NW}P_{w,t}^{s,\mathsf{ws}} - V_t^{\mathsf{PV}}\cdot\sum_{p=1}^{NP}P_{p,t}^{s,\mathsf{pvs}}\right\}
\end{aligned}
\tag{4.1}
$$

where $P_t$ is the hourly bid that the MG submits to the day-ahead market, $P_t^s$ is the real-time power delivery. The mismatch between the scheduled day-ahead power and the actual power delivery $|P_t^s-P_t|$ is indeed the amount of power that the MG trades in the real-time balancing market. A positive value of $P_t$ means that the power is exported from the MG to the main grid and vice versa. The same convention is applied to $P_t^s$ and $P_t^s-P_t$.

The proposed objective function represents the *expected profit* of the MG which is equal to the expected revenue attained by trading in both day-ahead and balancing markets minus the MG operating cost. The expected revenue of the MG is

$$
\sum_{s=1}^{NS}\rho_s\sum_{t=1}^{NH}\Delta T[P_t e_t^{s,\mathsf{DA}} + (P_t^s-P_t)e_t^{s,\mathsf{RT}} - \psi_t|P_t^s-P_t|].
\tag{4.2}
$$

In fact, if $P_t$ is positive (negative) then the term $\Delta T P_t e_t^{s,\mathsf{DA}}$ represents the revenue (cost) of the MG by selling (buying) electricity in the day-ahead market in time $t$. Similarly, if $(P_t^s-P_t)$ is positive (negative) then the term $\Delta T(P_t^s-P_t)e_t^{s,\mathsf{RT}}$ describes the revenue (cost) for the MG by selling (buying) electricity in the real-time market in time slot $t$ and scenario $s$. The term $\psi_t\Delta T|P_t^s-P_t|$ presents the penalty imposed on the MG aggregator in time





slot $t$ and scenario $s$ when the actual real-time power delivery is different from the scheduled day-ahead value. This revenue model is based on references [30, 32, 33, 49].

The MG operating cost consists of the startup cost, the shutdown cost, the operating cost of conventional units, and other costs including users' temperature discomfort cost, battery degradation cost, penalties due to wind/solar energy curtailment, and involuntary load curtailment. In particular, the total startup cost and the shutdown cost for conventional units over the scheduling horizon is expressed in the first term in (4.1) while the second term in (4.1) presents the operating cost of conventional units.

The third term in (4.1) represents the penalty due to temperature deviations in buildings. Specifically, $\pi_{j,t}|T_{j,t+1}^s - T_{j,t+1}^d|$ describes the temperature discomfort cost for residents of building $j$ in time slot $t$ and scenario $s$. The parameter $\pi_{j,t}$ represents the willingness of residents of building $j$ to trade their climate comfort for cost saving in time slot $t$. The larger $\pi_{j,t}$ is, the less willing the residents in building $j$ are, which results in less flexibility in scheduling power consumption of the HVAC system of building $j$. Here, $\pi_{j,t}|T_{j,t+1}^s - T_{j,t+1}^d|$ can be interpreted as the payment that the MG aggregator pays the residents of building $j$ in time slot $t$ and scenario $s$ for their participation in the underlying control scheme.

The term $C_k^{\deg}(\frac{P_{k,t}^{s,d}}{\eta_k^d} + \eta_k^c P_{k,t}^{s,c})$ captures the degradation cost for battery $k$ in time slot $t$ and scenario $s$ due to charging/discharging activities [36, 50, 106, 107]. In addition, the penalty for curtailment of involuntary load is proportional to the amount of load shedding. To ensure a high quality of service for users, involuntary load curtailment needs to be avoided, hence $V_t^{LL}$ should be set to a very high value. Finally, the last two terms in (4.1) represent the penalty for wind and solar energy curtailment, respectively. Renewable energy curtailment penalty is employed to account for the benefits associated with renewable energy that have not been included explicitly in the existing model (e.g., subsides from government to encourage increasing renewable energy penetration, carbon emissions reduction, and Renewable Energy Certificate (REC) policy) [99, 105]. We propose to include the renewable energy curtailment penalty in the objective function to increase the flexibility of the proposed model, which provides the MG aggregator the mechanism to efficiently control the amount of curtailment [1]. In general, the higher values of weighting factors $V_t^W$ and/or $V_t^{PV}$ results in a lower

---

[1]In many cases, penalty for renewable energy curtailment (or, also the discomfort cost) might not have actual economic values (e.g., there is no subside policies for renewable energy, no REC policy, etc or users do not receive the exact amount of discomfort cost as modeled in (4.1) ), we might still want to put the terms representing those penalties in the objective function with the purpose of allowing the MG aggregator more flexibly in controlling the amount of renewable energy curtailment, and users discomfort. However, the cost for those penalties will not be included in the real cost (or profit) value for the MG aggregator.





amount of renewable energy curtailment. Other constraints of the optimization problem are described in the following.

## 4.2.2 Power Balance

For each scenario $s$, the sum of the total power generation from all local generating units, the amount of involuntary load curtailment, and the charging/discharging power of battery units must be equal to the sum of the real-time power delivery, HVAC and non-HVAC loads. Note that $P_{w,t}^s$ and $P_{w,t}^{s,\text{ws}}$ are the maximum available wind power and the amount of wind power curtailment at time $t$ in scenario $s$, respectively. The difference between them is the actual wind power generation at time $t$ in scenario $s$. Similar explanation is applied to solar power generation. The power balance equation for each time $t$ and scenario $s$ is given as follows:

$$\sum_{i=1}^{NG} P_{i,t}^s + \sum_{w=1}^{NW} (P_{w,t}^s - P_{w,t}^{s,\text{ws}}) + \sum_{p=1}^{NP} (P_{p,t}^s - P_{p,t}^{s,\text{pvs}}) + LS_t^s + \sum_{k=1}^{NK} (P_{k,t}^{s,\text{d}} - P_{k,t}^{s,\text{c}})$$
$$= P_t^s + \sum_{j=1}^{NB} P_{j,t}^{s,\text{hvac}} + L_t^s, \quad \forall\, s,\, t. \quad (4.3)$$

## 4.2.3 Power Exchange with Main Grid

We can impose a limit on the quantity of power submitted to the day-ahead market as well as the amount of real-time power delivery. Let $P_t^{\text{g,max}}$ be the capacity of the line connecting the MG with the main grid then we have

$$-P_t^{\text{g,max}} \leq P_t^s \leq P_t^{\text{g,max}}, \quad \forall\, s,\, t. \quad (4.4)$$

$$-P_t^{\text{g,max}} \leq P_t \leq P_t^{\text{g,max}}, \quad \forall\, t. \quad (4.5)$$

## 4.2.4 Constraints for Conventional Units

The operating cost of conventional unit $i$ can be modeled approximately by a piecewise linear function as follows [99, 108]:

$$C(P_{i,t}^s) = a_i I_{i,t} + \Delta T \sum_{m=1}^{N_i} \lambda_{i,t}(m) P_{i,t}^s(m), \quad \forall\, i,\, t,\, s \quad (4.6)$$

$$0 \leq P_{i,t}^s(m) \leq P_{i,t}^{\text{max}}(m), \quad \forall\, i,\, t,\, s \quad (4.7)$$

$$P_{i,t}^s = P_i^{\text{min}} I_{i,t} + \sum_{m=1}^{N_i} P_{i,t}^s(m), \quad \forall\, i,\, t,\, s \quad (4.8)$$





where $N_i$ is number of segments of energy production curve for unit $i$ and $\lambda_{i,t}(m)$ is the marginal cost of the segment $m$ offered by unit $i$ in time slot $t$ ($/kWh) [99]. $a_i$ is the cost of running unit $i$ at its minimum power generation [108]. The following constraints represent the output power generation limits (4.9), ramping down/up rate limits (4.10)–(4.11), minimum ON/OFF time limits (4.12)–(4.13), and the relationship between the start-up and shutdown indicators ($y_{i,t}$ and $z_{i,t}$) (4.14)–(4.15) of the conventional generating unit $i$ [35, 83, 99, 100, 108]

$$P_i^{\min} I_{i,t} \leq P_{i,t}^s \leq P_i^{\max} I_{i,t}, \quad \forall\, i,\, t,\, s \tag{4.9}$$

$$P_{i,t}^s - P_{i,t-1}^s \leq UR_i(1 - y_{i,t}) + P_i^{\min} y_{i,t}, \quad \forall\, i,\, t,\, s \tag{4.10}$$

$$P_{i,t-1}^s - P_{i,t}^s \leq DR_i(1 - z_{i,t}) + P_i^{\min} z_{i,t}, \quad \forall\, i,\, t,\, s \tag{4.11}$$

$$\sum_{h=t}^{t+UT_i-1} I_{i,t} \geq UT_i y_{i,t}, \quad \forall\, i,\, t \tag{4.12}$$

$$\sum_{h=t}^{t+DT_i-1} (1 - I_{i,t}) \geq DT_k z_{i,t},, \quad \forall\, i,\, t \tag{4.13}$$

$$y_{i,t} - z_{i,t} = I_{i,t} - I_{i,t-1},, \quad \forall\, i,\, t \tag{4.14}$$

$$y_{i,t} + z_{i,t} \leq 1, \quad \forall\, i,\, t \tag{4.15}$$

$$I_{i,t} \in \{0,\, 1\},\, y_{i,t} \in \{0,\, 1\},\, z_{i,t} \in \{0,\, 1\}, \quad \forall\, i,\, t \tag{4.16}$$

Start-up cost and shut down cost constraints are given as follows [35, 36, 99, 108]:

$$SU_{i,t} \geq CU_{i,t}(I_{i,t} - I_{i,t-1}), \quad \forall\, i,\, t \tag{4.17}$$

$$SU_{i,t} \geq 0, \quad \forall\, i,\, t \tag{4.18}$$

$$SD_{i,t} \geq CD_{i,t}(I_{i,t-1} - I_{i,t}), \quad \forall\, i,\, t \tag{4.19}$$

$$SD_{i,t} \geq 0, \quad \forall\, i,\, t \tag{4.20}$$

Interested readers can find more details about modeling conventional units in [108].

### 4.2.5 Thermal Dynamic Model

A popularly used third-order state-space model that we studied in Section 2.1 is employed to describe the thermal dynamic model for buildings. This model captures the impacts of ambient temperature and solar irradiance on the indoor temperature [11, 40, 42].

$$T_{j,t+1}^s = A_j^{\mathsf{d}} T_{j,t}^s + B_j^{\mathsf{d}} U_{j,t}^s, \quad \forall j,\, t,\, s \tag{4.21}$$

$$T_{j,t}^{s,\text{in}} = C_j^{\mathsf{d}} T_{j,t}^s, \quad \forall j,\, t,\, s \tag{4.22}$$





where $T_{j,t}^s = [T_{j,t}^{s,in}, \ T_{j,t}^{s,im}, \ T_{j,t}^{s,om}]'$ is the state vector. $T_{j,t}^{s,im}$ is the temperature of the thermal accumulating layer in the inner walls and floor in building $j$ in time slot $t$ and scenario $s$ ($^oC$). $T_{j,t}^{s,om}$ is the temperature of the envelope of the of building $j$ in time slot $t$ and scenario $s$ ($^oC$). $U_{j,t}^s = [T^{s,a}, \ \Phi_t^s, \ \sigma_j \eta_j P_{j,t}^{s,hvac}]'$ is the input control vector. The coefficients of matrices $A_j^d$ and $B_j^d$ of building $j$ can be calculated based on the effective window area, the fraction of solar irradiation entering the inner walls and floor, and the thermal capacitance, thermal resistance parameters of the building and $C_j^d = [1, \ 0, \ 0]$. The constraints on the indoor temperature and HVAC power consumption are represented as follows:

$$T_{j,t}^d - \delta_{j,t} \leq T_{j,t}^{s,in} \leq T_{j,t}^d + \delta_{j,t}, \quad \forall j, \ t, \ s \tag{4.23}$$

$$0 \leq P_{j,t}^{s,hvac} \leq P_j^{hvac,max}, \quad \forall j, \ t, \ s \tag{4.24}$$

### 4.2.6 Battery Constraints

The constraints (4.25)–(4.28) capture the limits on the charging and discharging power as well as the level of energy stored in a battery unit $k$. Here, the level of battery storage at the end of the scheduling horizon is equal to its initial energy level. Constraints (4.29)-(4.30) are imposed to ensure the battery cannot be charged and discharged simultaneously in any time slot. The energy dynamic model for battery $k$ is captured in (4.31).

$$0 \leq P_{k,t}^{s,c} \leq b_{k,t}^{s,c} P_k^{c,max}, \quad \forall k, \ t, \ s \tag{4.25}$$

$$0 \leq P_{k,t}^{s,d} \leq b_{k,t}^{s,d} P_k^{d,max}, \quad \forall k, \ t, \ s \tag{4.26}$$

$$E_k^{min} \leq E_{k,t}^s \leq E_k^{max}, \quad \forall k, \ t, \ s \tag{4.27}$$

$$E_{k,NH}^s = E_{k,1}^s, \quad \forall k, \ t, \ s \tag{4.28}$$

$$b_{k,t}^{s,c} + b_{k,t}^{s,d} = 1, \quad \forall k, \ t, \ s \tag{4.29}$$

$$b_{k,t}^{s,c}, \ b_{k,t}^{s,d} \in \{0, \ 1\}, \quad \forall k, \ t, \ s \tag{4.30}$$

$$E_{k,t+1}^s = E_{k,t}^s + (\eta_k^c P_{k,t}^{s,c} \Delta T - \frac{P_{k,t}^{s,d} \Delta T}{\eta_k^d}), \quad \forall k, \ t, \ s. \tag{4.31}$$

### 4.2.7 Involuntary Load Curtailment

Constraints (4.32) limit the amount of involuntary load curtailment at time $t$. The expected amount of involuntary load curtailment at time $t$ is $\sum_{s=1}^{NS} \rho_s LS_t^s$, while the expected total non-HVAC load at time $t$ is $\sum_{s=1}^{NS} \rho_s L_t^s$. In this study, we force the expected involuntary load curtailment is smaller than a certain percentage of the expected non-HVAC load for





every time slot as described in (4.33). Constraints (4.33) can be considered as a reliability criterion for the operation of the MG.

$$0 \leq LS_t^s \leq LS_t^{\text{max}}, \quad \forall\ t,\ s \tag{4.32}$$

$$\frac{\sum_{s=1}^{NS} \rho_s LS_t^s}{\sum_{s=1}^{NS} \rho_s L_t^s} \leq LOL_t^{\text{max}}, \quad \forall t. \tag{4.33}$$

## 4.2.8 Renewable Energy Curtailment

The available output power of solar unit $p$ at maximum power point (MPP) in time slot $t$ and scenario $s$ can be calculated as follows [34]:

$$P_{p,t}^s = \eta_p S_p \Phi_t^s (1 - 0.005(T_t^{s,\text{a}} - 25)) \tag{4.34}$$

Further details on this solar power model are available in [109, 110]. The output power of wind generator $w$ in time slot $t$ and scenario $s$ is given as follows [34]:

$$P_{w,t}^s = \begin{cases} 0, & \text{if } v_t^s \leq v_w^{\text{ci}} \text{ or } v_t^s \geq v_w^{\text{co}}, \\[2mm] P_w^r \frac{v_t^s - v_w^{\text{ci}}}{v_w^r - v_w^{\text{ci}}}, & \text{if } v_w^{\text{ci}} \leq v_t^s \leq v_w^r, \\[2mm] P_w^r, & \text{otherwise.} \end{cases} \tag{4.35}$$

In each time slot, the amount of wind/solar power curtailment must obviously be smaller than the available wind/solar output power. Hence, we must have

$$0 \leq P_{w,t}^{s,\text{ws}} \leq P_{w,t}^s, \quad \forall\ w,\ t,\ s. \tag{4.36}$$

$$0 \leq P_{p,t}^{s,\text{pvs}} \leq P_{p,t}^s, \quad \forall\ p,\ t,\ s. \tag{4.37}$$

Note that spinning reserve and voluntary demand response load are not considered in our model; however, their integration into the model is straightforward.

## 4.2.9 Computation Time

The power scheduling and bidding problem for joint optimization of HVAC systems and distributed resources in the MG described in the previous section is a mixed integer linear program (MILP), which can be solved effectively by using available commercial solvers such as CPLEX [86]. The absolute terms in the objective function (4.1) can be easily transformed





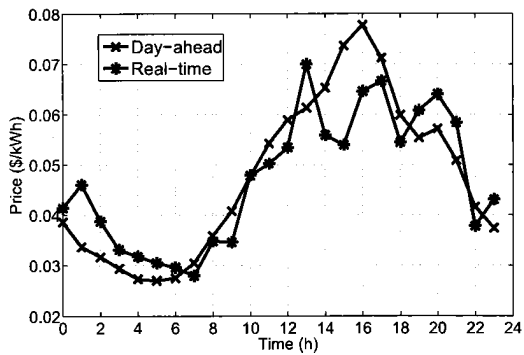

**Figure 4.1:** Hourly forecasted day-ahead and real-time electricity prices

into equivalent linear functions by introducing some auxiliary variables [91] as described in Section 2.2.2.

All the test cases presented in Section 4.3 are implemented on a desktop computer with 3.5 GHz Intel Core i7-3370 CPU and 16GB RAM. The computational time needed to run scenario reduction from 3000 scenarios to 15 scenarios using GAMS/SCENRED is recorded to be about 105 seconds. The calculation time (using CPLEX 12.4) for the proposed model with the 15 reduced scenarios is about 1 second, which is pretty small.

## 4.3 Numerical Results

We consider a MG whose portfolio consists of three conventional generating units, one wind turbine, one solar source, 100 buildings with their associated loads, and an optional battery facility. The parameters of three conventional units including two microturbines (MT) and one fuel cell (FC) are taken from [34], which are summarized in Table 4.1. For simplicity, the operating cost of each unit $i$ is modeled by a single curve segment (m = 1) [34, 99]. The shutdown cost is assumed to be 10% of the start-up cost. The parameter IC (i.e., Initial Condition) in Table 4.1 presents the number of hours that a unit is ON (positive) or OFF (negative).

We take the building thermal data from [41] and use the approach in Chapter 3 to model the diversity of thermal characteristics of buildings. We consider a summer case in this study; however, results for the winter case can be obtained similarly. We consider a 24-hour scheduling period where one time slot is one hour. Unless stated otherwise, we will set $\delta_{j,t} = \delta_T$, $\forall j, t$; $\pi_{j,t} = \pi$, $\forall j, t$; $V_t^{PV} = V_t^W = V^{RES}$, $\forall t$; and $\psi_t = \psi$, $\forall t$.





**Table 4.1:** Conventional unit data

| Gen # | Type | a ($) | $\lambda(\$/kWh)$ | $P^{min}(kW)$ | $P^{max}(kW)$ |
|---|---|---|---|---|---|
| 1 | MT | 30 | 0.13 | 100 | 2000 |
| 2 | MT | 50 | 0.35 | 100 | 1000 |
| 3 | FC | 80 | 0.5 | 100 | 1000 |

| Gen # | CU ($) | CD ($) | UT (hrs) | DT (hrs) | IC (hrs) |
|---|---|---|---|---|---|
| 1 | 150 | 15 | 2 | 2 | -2 |
| 2 | 30 | 3 | 0 | 0 | -1 |
| 3 | 30 | 3 | 0 | 0 | -1 |

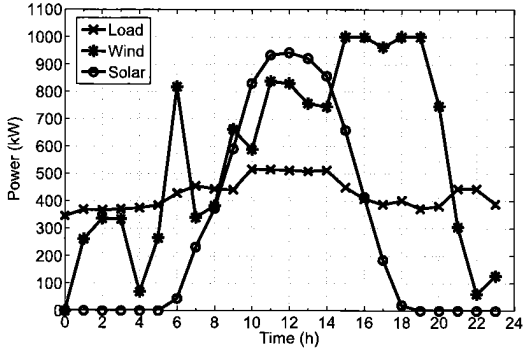

(a) Wind, solar and non-HVAC load forecasts

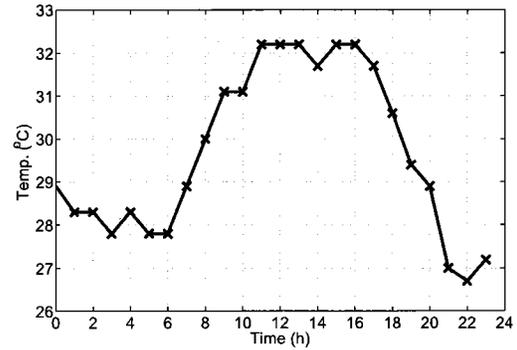

(b) Hourly forecasted ambient temperature

**Figure 4.2:** Wind power, solar power, non-HVAC load, and temperature forecasts

The forecasts for uncertain parameters in the system model are assumed to be available. To run the simulation, we use the historical data for wind speed [111], non-HVAC load [100], solar irradiance [94], ambient temperature [93], day-ahead and real-time electricity prices [95] with appropriate scaling coefficients as in the forecasts. Figs. 4.1, 4.2(a), and 4.2(b) show the hourly forecasts for the uncertain parameters in the considered model. Wind output power and solar output power can be calculated from the wind speed, solar irradiance, and ambient temperature by using (4.34) and (4.35), respectively. The parameters of the wind





and solar sources are retrieved from [34] as follows:

- For wind source: $P^r = 1000$ kW, $v^{ci} = 3$ m/s, $v^r = 12$ m/s, $v^{co} = 30$ m/s.

- For solar source: $\eta = 15.7\%$ and S = 7000 m$^2$. Under standard condition test (SCT) with ambient temperature of 25°C, solar irradiance of 1000 W/m$^2$, and at the maximum power point (MPP) [109, 110], the rated PV power is 1100 kW.

System parameters for *the base case* are set as follows. The value of lost load $V_t^{LL}$ is set to 1000 \$/MWh, the bid deviation penalty cost $\psi_t$ is set to 80 \$/MWh, and no penalty cost for renewable power generation curtailment and indoor temperature deviation. The maximum involuntary load curtailment is set equal to 5% of the expected non-HVAC load in each time slot and no battery storage unit is included. Also, we do not consider the maximum power exchange constraints (4.5). Note that when we do not set a limit on the amount of power submitted to the day-ahead market, we need to set $\psi_t$ sufficiently high to ensure the bid deviation is not too large and the actual power delivery is close the the day-ahead schedule.

We define the Load Scaling Factor (LSF) as the ratio between the total forecasted non-HVAC load and the total forecasted renewable generation over the scheduling period. For example, the LSF in Fig. 4.2(a) is equal to 0.5, which is chosen in the base case. For simplicity, we assume that all buildings are always occupied over the scheduling period. The desired temperature is set to 23°C and the maximum allowable temperature deviation is set to 2°C for all buildings in any time slots. The initial indoor temperatures of all buildings are assumed to be equal to the desired indoor temperature 23°C. HVAC rated power in all buildings is set to 10 kW. The data defining *the base case* is given in Table 4.2. For all figures presented in this section, only the system parameters explicitly presented in the figures are varied, other parameters are the same as in the base case.

Three control schemes are studied in this section as follows:

**Scheme 1**: In this scheme, we apply to the proposed optimal scheme to the considered MG.

**Scheme 2**: In this scheme, we still apply the proposed optimal control scheme to the MG; however, the indoor temperatures of buildings are always maintained at the setpoint. In other word, no temperature deviation is allowed ($\delta_T = 0$).

**Scheme 3** (uncoordinated optimal scheme): In this scheme, HVAC systems and the rest of the MG optimize their power profiles separately (Problems 1 and 2 described below). Here, the objective of HVAC scheduling is set to minimize the operation cost of HVAC systems. HVAC systems submit their aggregated demand bids to the day-ahead market and face the same penalty scheme for bid deviation charge as we described above. The





**Table 4.2:** System parameters in the base case

| $V_t^{LL}$ ($/MWh) | $\psi$ ($/MWh) | $V^{RES}$ ($/MWh) | $\pi$ ($/^oC$) |
|---|---|---|---|
| 1000 | 80 | 0 | 0 |

| $\delta_T$ ($^oC$) | LSF | $P_j^{\text{hvac,max}}$ (kW) | Battery |
|---|---|---|---|
| 2 | 0.5 | 10 | No |

technical constraints for HVAC system, thermal constraints for buildings, and others remain the same. The only the difference is that the power balancing equation (4.38) does not have the HVAC-related terms.

**Problem 1:** We aim to minimize the expected operating cost of HVAC systems as follows:

$$\min \quad \sum_{s=1}^{NS} \rho_s \sum_{t=1}^{NH} \Delta T \left\{ P_t^{\text{hvac}} e_t^{s,\text{DA}} + (P_t^{s,\text{hvac}} - P_t^{\text{hvac}}) e_t^{s,\text{RT}} \right.$$

$$\left. + \pi_{j,t} \sum_{j=1}^{NB} |T_{j,t+1}^s - T_{j,t+1}^d| + \psi_t |P_t^{s,\text{hvac}} - P_t^{\text{hvac}}| \right\}$$

$$\text{s. t.} \quad P_t^{s,\text{hvac}} = \sum_{j=1}^{NB} P_{j,t}^{s,\text{hvac}}, \forall t, \ s$$

and other constraints for HVAC system and thermal comfort requirement (4.21), (4.22), (4.23), and (4.24). Here, $P_t^{\text{hvac}}$ and $P_t^{s,\text{hvac}}$ are positive numbers which denote the imported power for HVAC systems at time $t$.

**Problem 2:** For this problem, the expected profit for the MG is maximized. The objective function remains the same as (4.1) but the discomfort cost term is not included. The power balance equation now becomes

$$\sum_{i=1}^{NG} P_{i,t}^s + \sum_{w=1}^{NW} (P_{w,t}^s - P_{w,t}^{s,\text{ws}}) + \sum_{p=1}^{NP} (P_{p,t}^s - P_{p,t}^{s,\text{pvs}}) + \sum_{k=1}^{NK} (P_{k,t}^{s,\text{d}} - P_{k,t}^{s,\text{c}}) + LS_t^s = P_t^s + L_t^s, \forall t, \ s$$

and other constraints for the components in the MG are unchanged as described in the problem formulation.

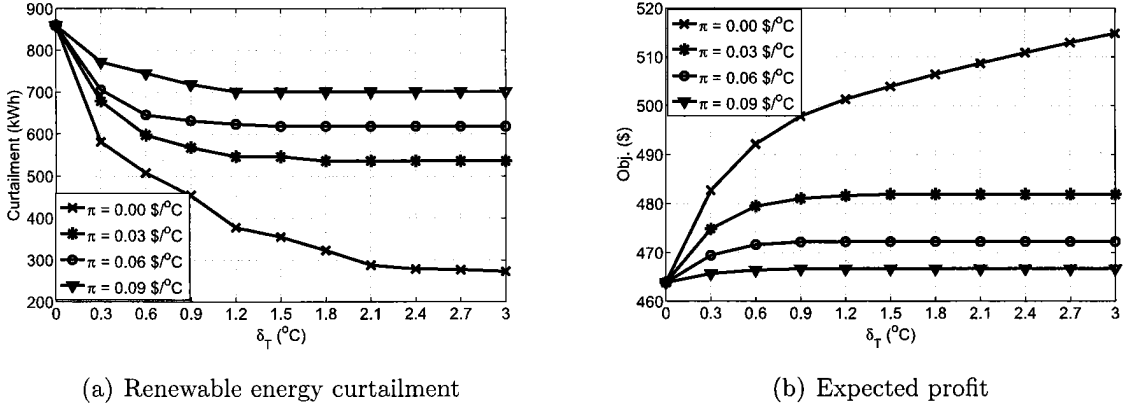

(a) Renewable energy curtailment          (b) Expected profit

**Figure 4.3:** Impacts of $\delta_T$ and $\pi$ on the optimal solution

## 4.3.1 Comparison Between Scheme 1 and Scheme 2

Figs. 4.3(a) and 4.3(b) illustrate the advantages of the proposed optimal scheme, which exploits the flexibility offered by building thermal storage capacity, and the flexible comfort requirements (Scheme 1) compared to the case where no temperature deviation is allowed (Scheme 2 with $\delta_T = 0$). As discussed in Section 4.1.1.3, the curtailment of renewable energy generation is required in some cases to avoid the high penalty charge on bid deviation and to ensure that the real time power delivery is sufficiently close to the day-ahead schedule. Due to their flexibility, the HVAC systems could increase their power consumption by absorbing more energy from renewable sources to help the MG reduce the amount of renewable energy curtailment.

Fig. 4.3(a) confirms that the amount of renewable energy curtailment is reduced as $\delta_T$ increases. Furthermore, the increase in discomfort penalty cost $\pi$ results in more curtailed renewable energy. Fig. 4.3(b) shows the expected profit (i.e., the value of the objective function) for different values of $\delta_T > 0$ and $\pi$. As evident, as $\delta_T$ increases, the MG expected profit increases. Also, as the cost of temperature deviation $\pi$ increases, the MG expected profit decreases. In fact, larger values of $\delta_T$ (and/or smaller values of $\pi$) enables more flexible scheduling of the HVAC power consumption, which results in larger performance gain in terms of both MG profit and renewable energy curtailment reduction.

## 4.3.2 Comparison Between Scheme 1 and Scheme 3

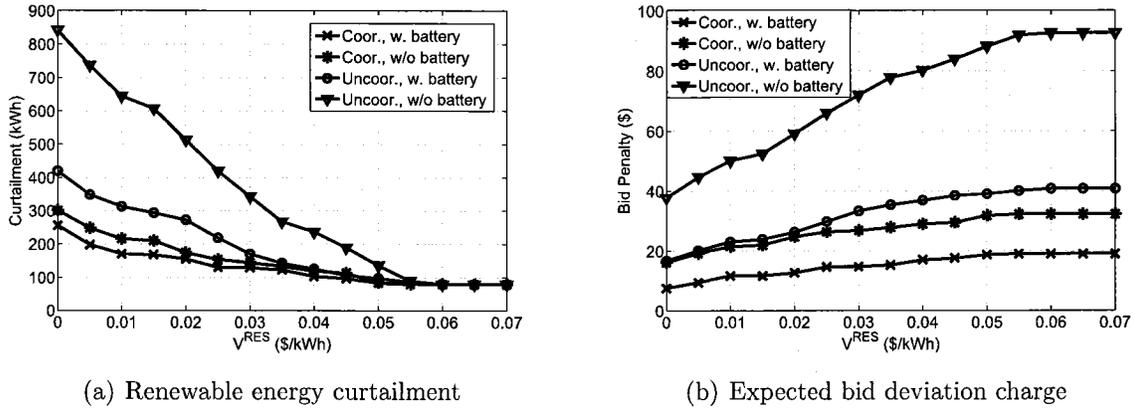

(a) Renewable energy curtailment        (b) Expected bid deviation charge

**Figure 4.4:** Comparison between coordinated and uncoordinated schemes

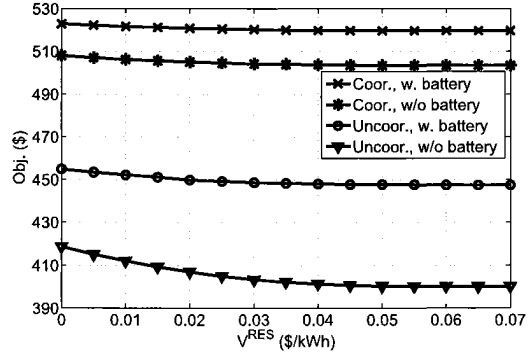

**Figure 4.5:** Expected profit in coordinated and uncoordinated schemes

Figs. 4.4(a), 4.4(b), and 4.5 show the advantages of the coordinated optimal scheme compared to the uncoordinated scheme versus the curtailment parameter $V^{RES}$. We compare the two schemes in two cases with and without battery. For the case with battery, one battery unit with capacity of 200 kWh is chosen where the charging/discharging power ratings are set equal to 100 kW. The minimum and maximum energy stored in the battery are 40 kWh and 180 kWh, respectively. Table 4.3 summarizes the parameters of the considered battery unit. We can see that the total amount of renewable energy curtailment is much smaller in the coordinated scheme compared with that of the uncoordinated one. This is because in the uncoordinated scheme, we do not exploit the flexibility offered by HVAC systems to absorb the fluctuation of renewable energy generation.

To mitigate the high penalty due to the bid deviation charge, some surplus renewable energy needs to be curtailed to ensure that the real-time power delivery is close to the day-ahead schedule. However, in the coordinated scheme, HVAC systems can increase their power consumption when the renewable energy sources produce surplus energy, which helps reduce the amount of renewable energy curtailment. Additionally, it can be observed that the amount of renewable energy curtailment decreases as $V^{RES}$ increases and it tends to zeros as $V^{RES}$ becomes larger. Fig. 4.4(b) shows that the bid deviation charges for the uncoordinated scheme is much higher than that in the coordinated scheme. This is because when the renewable energy generation is smaller than expected, in the coordinated scheme, we can reduce the power consumption of HVAC systems to reduce the charge due to the shortage of delivered power to the market as being scheduled in advance. Finally, the expected profit for the MG is also higher in the coordinated scheme than the one in the uncoordinated scheme as shown in Fig. 4.5.

**Table 4.3:** Battery parameters

| $E^{cap}$ (kWh) | $E^{min}$ (kWh) | $E^{max}$ (kWh) |
|---|---|---|
| 200 | 40 | 180 |

| $P^{c,max}$ (kW) | $P^{d,max}$ (kW) | $C^{wear}$ ($/kWh) |
|---|---|---|
| 100 | 100 | 0.00027[50] |

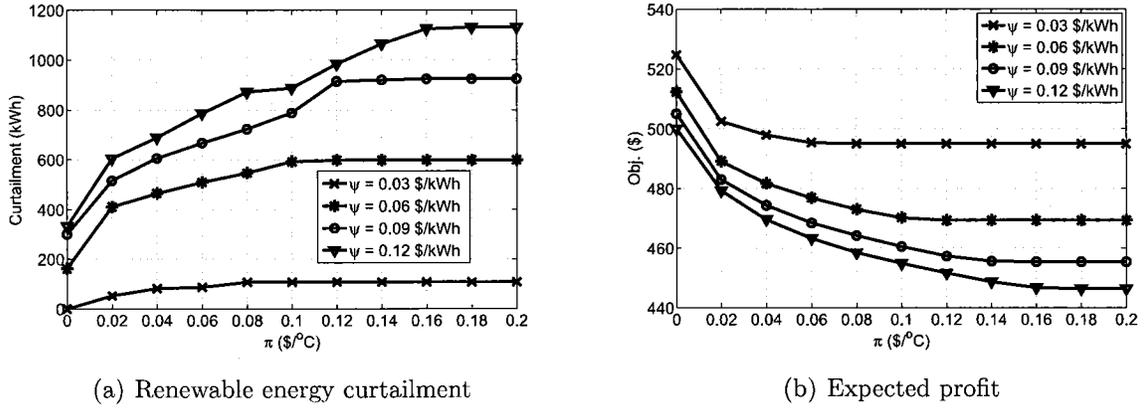

(a) Renewable energy curtailment

(b) Expected profit

**Figure 4.6:** Impact of temperature deviation penalty ($\pi$) on the optimal solution

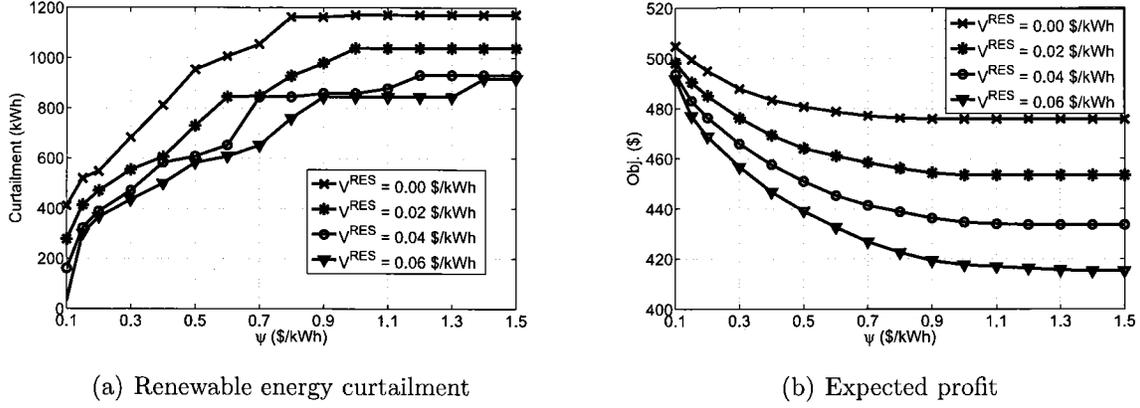

(a) Renewable energy curtailment       (b) Expected profit

**Figure 4.7:** Impacts of $\psi$ and $V^{RES}$ on the optimal solution

## 4.3.3 Sensitivity Analysis

We now investigate the impacts of various design and system parameters on the optimal solution. Figs. 4.6(a) and 4.6(b) illustrate the impacts of temperature deviation penalty cost $\pi$ on the optimal solution. Specifically, Fig. 4.6(a) confirms that the energy curtailment tends to increase as $\pi$ increases. This is due to the fact that as $\pi$ increases, the indoor temperature is forced to be closer to the desired value ($T^d$) to reduce the climate discomfort cost, which means we have the less flexibility in controlling the HVAC power consumption. Moreover, the amount of renewable energy curtailment becomes saturated at certain values of $\pi$ where the indoor temperature of all buildings becomes very close to the desired temperature. Also, we can see that the expected profit of the MG decreases and becomes flattened as $\pi$ is sufficiently large as shown in Fig. 4.6(b).

In addition, as parameter $\psi$ increases, we expect that the bid deviation becomes smaller to avoid the high bid deviation penalty cost, which results in less flexibility in controlling the operation of the MG. Therefore, the expected profit for the MG decreases and the amount of renewable energy curtailment increases as $\psi$ increases, which is confirmed by the results in Figs. 4.6(a), 4.6(b), 4.7(a), and 4.7(b).

Figs. 4.8(a) and 4.8(b) illustrate the variation of the amount of total renewable energy curtailment and the MG expected profit versus the HVAC rated power. This figure shows that lower amount of renewable energy curtailment and higher MG expected profit can be achieved as the HVAC rated power increases, which indeed offers more flexibility in controlling the HVAC power consumption. However, with each value of battery capacity, the quantities are saturated at certain values of HVAC rated power, which can be interpreted

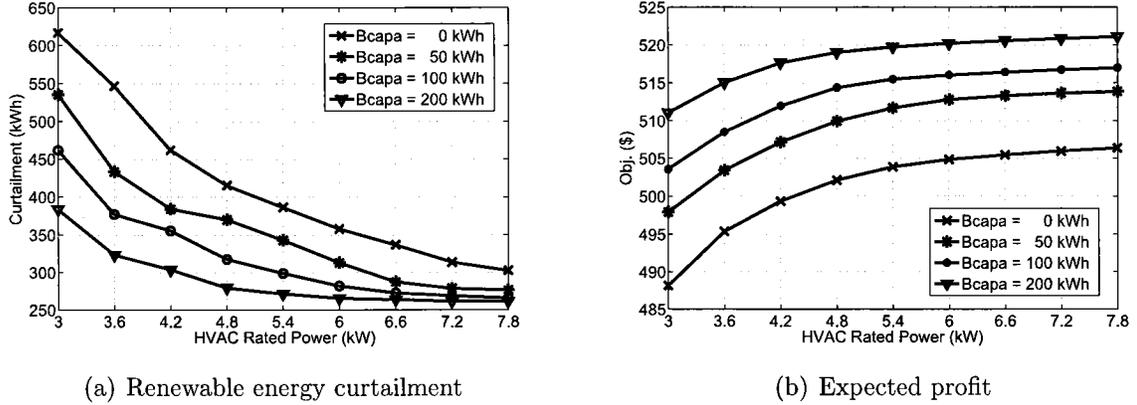

(a) Renewable energy curtailment

(b) Expected profit

**Figure 4.8:** Impacts of $\psi$ and $V^{RES}$ on the optimal solution

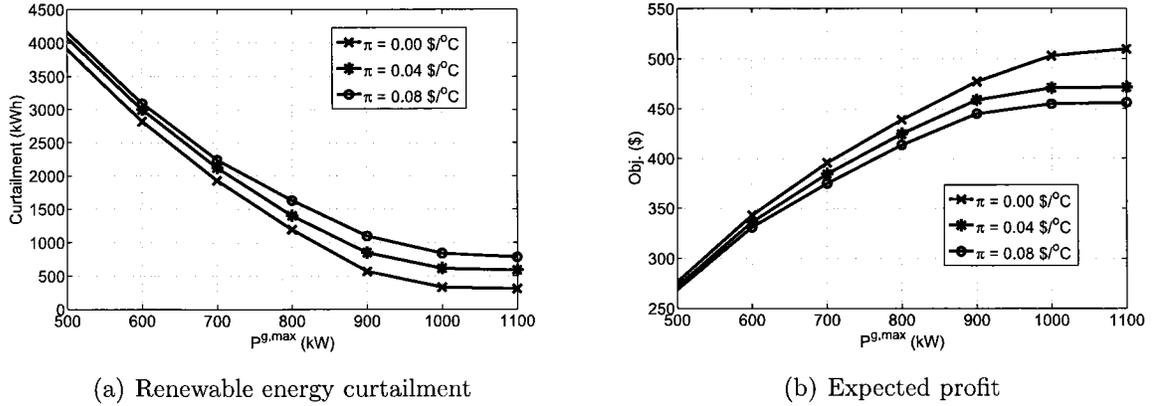

(a) Renewable energy curtailment

(b) Expected profit

**Figure 4.9:** Impact of $P^{g,max}$ on the optimal solution

as follows. The HVAC power consumption can only be varied within a certain range to meet the indoor temperature comfort requirement. Here, the minimum and maximum energy levels stored in the battery are set equal to 20% and 90% of the battery capacity. The maximum charging/discharging rate of battery is set equal to 100 kW.

The impact of the maximum allowable power exchange between the MG and the main grid ($P^{g,max}$) on the optimal solution is presented in Figs. 4.9(a) and 4.9(b). In particular, Fig. 4.9(a) shows that the amount of renewable energy curtailment increases for decreasing $P^{g,max}$. This can be interpreted as follows. If $P^{g,max}$ is small and the amount of available renewable energy is large, then the HVAC systems may not be able to absorb all the surplus renewable energy due to the temperature comfort constraint, and $P^{g,max}$ directly affects the ability of exchange power between the MG and the main grid. Therefore, more power curtailment is expected as $P^{g,max}$ is small. Also, the $P^{g,max}$ parameter has a direct impact on the ability of the MG in offering bid to the power market; therefore, the expected profit of the MG increases before getting saturated as $P^{g,max}$ increases as being shown in Fig. 4.9(b).

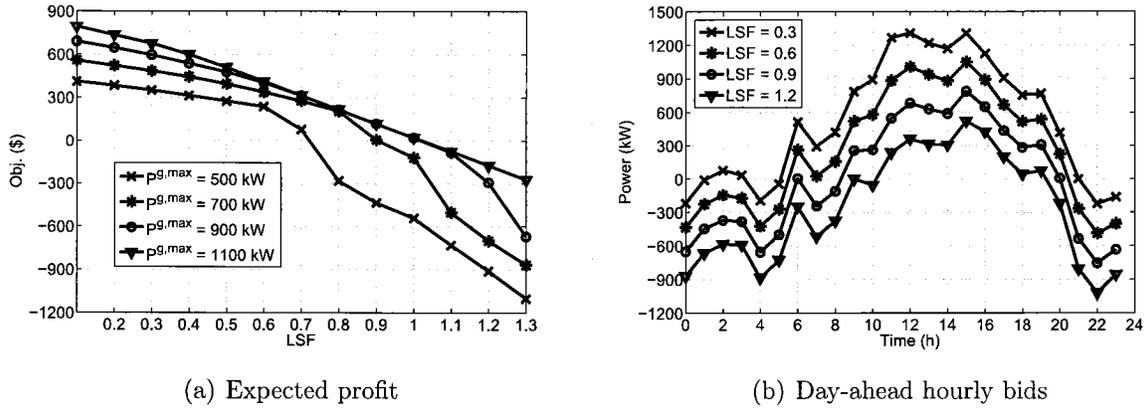

(a) Expected profit

(b) Day-ahead hourly bids

**Figure 4.10:** Impacts of $P^{g,max}$ and LSF on the optimal solution

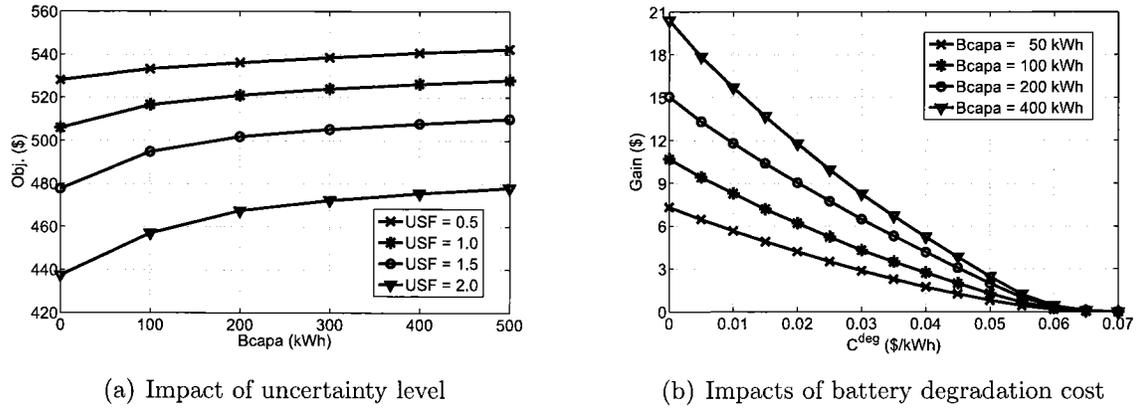

(a) Impact of uncertainty level

(b) Impacts of battery degradation cost

**Figure 4.11:** Impacts of uncertainty level and battery degradation cost on the optimal solution

The variations in expected profit of the MG with different values of the LSF and $P^{g,max}$ are presented in Fig. 4.10(a). Here, the negative value of expected profit corresponds to the case where the MG needs to purchase additional energy from the main grid on average; in other words, the MG has to pay the main grid operator for its operation. This is the case when the local load consumes more energy than what can be generated by the local renewable energy sources. The deficit amount of energy must, therefore, be compensated by importing energy from the main grid. Note also that as the LSF is sufficiently high, the MG may need to operate its local conventional generation units due to the constraints on the maximum amount of imported power. The conventional units in the considered model are only needed for backup purposes because their operation cost is relatively high, which implies that importing energy from the grid might be more cost-efficient than running conventional units to serve the local demand. The optimal bid quantities submitted to the day-ahead market for different values of the LSF are shown in Fig. 4.10(b). Here, if the power bid is negative, the MG imports power from the main grid.

To investigate the impact of the uncertainty level on the optimal solution, we utilize the *uncertainty scaling factor* (USF) where *the base case* described in Section 4.1.2 has USF = 1. Also, we scale the standard deviations of uncertain parameters in the base case by the factor of USF to obtain the result presented in Fig. 4.11(a). It can be observed that the MG expected profit decreases as USF increases (i.e., the uncertainty level increases).

Finally, we show the impact of battery operation cost (degradation) parameter $C^{\text{deg}}$ on the gain as battery storage is utilized. The presented gain captures the difference between the expected profit when using battery and when not using the battery. It can be observed from Fig. 4.11(b) that the gain due to utilizing battery storage facility decreases and becomes saturated as $C^{\text{deg}}$ increases. This can be interpreted as follows. If $C^{\text{deg}}$ is sufficiently high, the cost saving due to energy storage can be neutralized by the battery degradation (wear) cost. In general, if $C^{\text{deg}}$ is small then utilization of battery storage can result in positive performance gain.

## 4.4 Summary

In this chapter, we presented an optimal power scheduling for the MG with renewable energy considering users' thermal comfort requirements, day-ahead pricing, and various system constraints. The power scheduling and bidding problem was formulated as a two-stage stochastic program where we aim to maximize the expected profit of the MG in the deregulated electricity market. The thermal storage capability of buildings is exploited to counteract the uncertainties due to intermittent renewable energy sources. Extensive numerical results were presented to illustrate the great benefits for the MG in reducing the renewable energy curtailment, mitigating the high penalty resulted from the energy imbalance, and increasing the expected profit for the MG through exploiting the building thermal dynamics.

# Chapter 5

# Conclusions and Future Work

## 5.1 Conclusion Remarks

This thesis focused on home and building energy scheduling design issues where consideration of thermal dynamics with user comfort preferences and HVAC control are in the center of the study. In general, HVAC power consumption is controlled to track the desired temperature setpoint, which is set by users (e.g, via a thermostat). However, by considering flexible user comfort ranges and other system dynamics such as electricity price, weather conditions, and occupancy patterns, the HVAC load can be scheduled to minimize the electric bills for end users. In particular, HVAC load can be shifted to low-priced periods under the time-varying electricity pricing where the HVAC system might consume more power during low-priced periods to precool (preheat) the building in the summer (winter). Consequently, HVAC power consumption can be reduced during high-priced periods while maintaining temperature within the user temperature comfort range thanks to building thermal inertia. Since the penetration of renewable energy is still moderate in most current power systems, electricity price and power demand are highly correlated. As a result, high-priced periods are generally on-peak periods and low-priced periods generally coincident with off-peak periods. Therefore, smart scheduling of the HVAC load could not only reduce electricity cost but also reduce the peak power demand, which implies great social benefit.

In this thesis, the benefits of HVAC smart power scheduling was evaluated through two application scenarios. In the first application, we consider the joint scheduling optimization for HVACs and EVs. This design is motivated by the fact that they are expected to be among the most power-hungry appliances in the near future. Furthermore, HVAC and EV loads are flexible loads. Numerical studies were carried out for the developed solutions in two scenarios, namely single household and multiple households (e.g., a community setting). These studies showed that the performance of our proposed optimal solution is influenced significantly by various system parameters including travel patterns of EVs and the temperature comfort settings of users (i.e., maximum allowable temperature deviation and the cost of temperature deviation). Moreover, V2G can bring considerable benefits compared to the case where V2G is not allowed. Finally, if multiple households in a community jointly optimize the energy consumption of their EVs and HVACs then significant better performance can be achieved compared to the case where each household optimizes the energy consumption of their EV and HVAC independently.

In the second application, we considered the energy scheduling and bidding design for building microgrids where time-varying electricity price and intermittent renewable energy generation were modeled. Moreover, detailed operational aspects for various components in a renewable-powered MG were modeled. The flexible HVAC load was exploited to compensate for uncertainties in the DA energy bidding decision of the MG aggregator. Then, an optimal power scheduling framework for the MG with renewable energy considering user thermal comfort requirements and other system constraints was developed. Extensive numerical results were presented to illustrate the great benefits of our design in reducing the renewable energy curtailment, mitigating the high penalty due to energy imbalance, and increasing the expected profit for the MG by exploiting the building thermal dynamics.

The numerical studies revealed several interesting results, which are summarized as follows. First, coordination of HVAC load (or in general flexible loads) and RESs through a unified energy management framework can indeed result in significant increase in the profit of the MG aggregator. Second, the benefits of the proposed coordinated scheme depend on the flexibility offered by the HVAC system. Specifically, the expected profit of the MG aggregator increases and the amount of renewable energy curtailment decreases as the maximum allowable temperature deviation increases, the cost of temperature deviation decreases, and the rated power of HVAC systems increases. Third, the costs of bid deviation and renewable energy curtailment have significant impacts on the optimal solution. In particular, the MG expected profit decreases and the amount of renewable energy curtailment increases as the costs of bid deviation increases and/or renewable energy curtailment increases. Fourth, battery storage can help the MG aggregator reduce the amount of renewable energy curtailment and increase its expected profit. Fifth, the expected profit of the MG aggregator increases and the amount renewable energy curtailment decreases as the maximum power exchange limit between the MG aggregator and the main grid increases. However, it is saturated as this maximum power limit becomes sufficiently large. Finally, the uncertainty level has significant impacts on the optimal solution.

## 5.2  Future Research and Extensions

Although optimal power scheduling design for HVAC systems in two different application scenarios have been addressed in this thesis, many open problems remain to be answered in the design of the future smart grids where more active participation of demand side resources is expected. In the following, I would like to point out some potential research problems for further studies.

- In both applications that were considered in this thesis, the energy scheduling problems were formulated and solved in a centralized manner where a centralized aggregator is responsible for calculating the optimal solution for all buildings in the considering community or microgrid. However, it would be interesting to investigate the case where the MG aggregator is a *for-profit* entity that sells electricity to customers and might receive flexible loads offers from customers. Under this setting, the problem formulation would need to be revised since the objective of the controller is different, which is not simply to minimize the total energy cost of the community or to maximize the profit of the controller in the energy market. The aggregator might want to maximize the retail revenue of selling electricity to customers and minimize the cost of energy procurement in the energy market, operation cost of local distributed generators as well as the payment to customers for their flexible load offers.

- The cost of temperature deviation, which is the indicator of user discomfort cost, was chosen as a fixed and predefined value. How to design this parameter effectively is an interesting question since it represents the conflict of interests between the aggregator and customers. From the aggregator's perspective, increasing this cost implies less flexibility in controlling the HVAC power consumption. On the other hand, a small value of this cost would result in larger temperature deviation for the same penalty of discomfort, which affects user comfort.

- In the deregulated electricity market, demand side resources can participate in the ancillary service market (e.g., frequency regulation) as well as in the DA energy market (e.g., by offering its load reduction capacity bid). There are several references which have considered the ability of utilizing thermal load for frequency regulation service, or load following service. However, economic models and motivation in these research works are not clear. Furthermore, most of the existing works do not consider the effect of regulation signals on the HVAC power consumption which directly impacts the indoor temperature. Modeling the effects of regulation signal on the indoor temperature is challenging since the regulation signals can be highly uncertain. The inter-temporal dependence of temperatures in the building thermal dynamic model makes the problem even more complicated since the change of power consumption at a particular scheduling period will affect the temperature profile in the following periods. Therefore, finding the optimal offer capacity that a set of HVAC systems offer to the ancillary market is an interesting problem.

- In addition to the economic objectives of the MG aggregator (or a community) that we considered throughout this thesis, future works may consider other objectives that a MG aggregator may be interested in. For example, some utilities in the North America [112, 113] apply very high demand charges (i.e., the highest power consumption (kW) over a certain period) to their electric customers. Hence, a MG aggregator may be interested in minimizing their maximum power demand over a scheduling horizon, which consequently reduce the electric bills that the MG aggregator has to pay. This objective function is meaningful even when time-varying electricity pricing is not considered.

- Future works will quantify the savings for residential households over a certain time period (e.g., one year) if they improve the COP of their HVAC systems or they improve the insulation of their buildings.

In general, solving these open problems require innovations in both modeling and solution aspects. Therefore, much more research need to be performed in the next several years to fully understand and realize the potential benefits of the underlying designs that possess various interesting but complicated inter-component interactions.

## 5.3   List of Publication

1. **D.T. Nguyen** and L.B. Le, "Optimal bidding strategy for microgrids considering renewable energy and building thermal dynamics ," *IEEE Transaction on Smart Grid*, to be published.

2. **D.T. Nguyen** and L.B. Le, "Joint optimization of electric vehicle and home energy scheduling considering user comfort preference," *IEEE Transaction on Smart Grid*, vol. 5,

no. 1, pp. 188–199, Jan. 2014.

3. **D.T. Nguyen** and L.B. Le, " Optimal energy trading for building microgrid with electric vehicles and renewable energy resources," in *Proc. IEEE PES Innovative Smart Grid Technologies Conference (ISGT'14)*, Washington, DC, to appear.

4. Hieu. T. Nguyen, **D.T. Nguyen** and L.B. Le, "Home energy management with generic thermal dynamics and user temperature preference," in *Proc. of the IEEE Conference on Smart Grid Communications (SmartGridComm'13)*, Vancouver, Canada, Oct. 2013.

5. **D.T. Nguyen** and L.B. Le, "Optimal energy management for cooperative microgrids with renewable energy resources," in *Proc. of the IEEE Conference on Smart Grid Communications (SmartGridComm'13)*, Vancouver, Canada, Oct. 2013.